\documentclass[11pt,a4paper]{article}

\usepackage{jheppub}

\usepackage{graphics}
\usepackage{graphicx}
\usepackage{amsmath,amssymb,mathrsfs}
\usepackage{epsf}
\usepackage{slashed}
\usepackage{epsfig}
\usepackage{xcolor}
\allowdisplaybreaks

\newcounter{fig}

\newcommand{\bea}{\begin{eqnarray}}
\newcommand{\eea}{\end{eqnarray}}
\newcommand{\be}{\begin{equation}}
\newcommand{\ee}{\end{equation}}

\newcommand{\re}[1]{(\ref{#1})}

\newcommand{\eqn}{\begin{eqnarray}}
\newcommand{\eqnx}{\end{eqnarray}}

\begin{document}

\title{Resonance structures in kink-antikink collisions in a deformed sine-Gordon model}
\author[a]{Patrick Dorey,}
\author[b]{Anastasia Gorina,}
\author[b]{Ilya Perapechka,}
\author[c]{Tomasz Roma\'nczukiewicz}
\author[d]{and Yakov Shnir}

\affiliation[a]{Department of Mathematical Sciences, Durham University, UK}
\affiliation[b]{Department of Theoretical Physics and Astrophysics, Belarusian State University, Minsk 220004, Belarus}
\affiliation[c]{Faculty of Physics, Astronomy and Applied Computer Science, Jagiellonian University, Krak\'ow, Poland}
\affiliation[d]{BLTP, JINR, Dubna 141980, Moscow Region, Russia}

\emailAdd{p.e.dorey@durham.ac.uk}
\emailAdd{nastya.gorina.2931@gmail.com}
\emailAdd{jonahex111@outlook.com}
\emailAdd{tomasz.romanczukiewicz@uj.edu.pl}
\emailAdd{shnir@theor.jinr.ru}

\abstract{We study kink-antikink collisions in a
model which interpolates smoothly between the completely integrable
sine-Gordon theory, the $\phi^4$ model, and a $\phi^6$-like model with three degenerate vacua. We find a 
rich variety of behaviours, including integrability breaking, resonance windows with increasingly irregular patterns, and new types of windows near the $\phi^6$-like regime. False vacua, extra kink modes and kink fragmentation play important roles in the explanations of these phenomena. Our numerical studies are backed up by detailed analytical considerations.
}

\arxivnumber{XXXX.XXXX}

\maketitle

\section{Introduction}
The static and dynamic behaviour of kinks, that is topological solitons in
1+1 dimensional classical scalar field theories,
has attracted much attention over the years;
see e.g.\ \cite{Manton:2004tk,Vachaspati,Shnir2018}.
In the classical sine-Gordon  theory, the complete integrability of the field
equations leads to constraints on the dynamics of these solitons via the
infinite number of integrals of motion, as reviewed in \cite{babelon}.
As a result, the collision of sine-Gordon solitons cannot excite
radiation modes, and the asymptotic
structure of the final configuration only differs
from what would have been found in the absence of any interaction by a set of phase shifts. 

However, integrable models are exceptional, and most physical theories do not belong to this
class. Usually, an integrable model can be considered as an approximation to a more
realistic theory, and while some features of the integrable model may still persist in a deformed non-integrable
system, others are lost, often in interesting ways. Aspects of this interplay between integrable and quasi-integrable models have attracted
a lot of attention recently, see e.g.\ \cite{Ferreira:2012sd,Ferreira:2010gh,Uiimo}.
One example is the modified sine-Gordon model of \cite{Peyrard1,Peyrard2} which
still
supports kink solutions, though their scattering is much more complicated
than in the original integrable  model. The presence of defects (or
impurities) and boundaries can also destroy the complete integrability of the sine-Gordon theory in a variety of ways \cite{malomed,malomed2,Arthur:2015mva}.

The $\phi^4$ theory is similar to the
sine-Gordon model in that they both support topological kinks, but
the $\phi^4$ theory is not exactly integrable, and there is radiative energy loss
in kink-antikink ($K\bar K$) collisions.  Associated with this is the resonant energy exchange mechanism and the resulting
chaotic dynamics of kinks, see e.g.\
\cite{Anninos:1991un,Campbell:1983xu,Goodman:2005,Makhankov:1978rg,Moshir:1981ja,Dorey:2011yw}.
Numerical simulations of the $\phi^4$ theory reveal that for certain initial
velocities, the kink and antikink collide, become separated by a finite distance,
then collide a second time before finally escaping to infinity.
This is known as a (two-bounce) resonance window \cite{Anninos:1991un,Campbell:1983xu}.
Solitons can also escape after three or more consecutive collisions, leading to
an intricate nested structure of multi-bounce windows \cite{Anninos:1991un,Campbell:1983xu}.
These resonance windows are related
to the reversible exchange of energy between the translational
mode of the $\phi^4$ kink and its internal mode \cite{Anninos:1991un,Campbell:1983xu}.
This mechanism can be reasonably well approximated in a truncated model, which takes
into account only two dynamical degrees of freedom, the collective coordinates of the kink and
the internal mode \cite{Anninos:1991un,Goodman:2005,Takyi:2016tnc,Weigel:2013kwa}. 
This approach is qualitatively effective but many attempts to derive it from first principles have been plagued by many ambiguities and discrepancies, see \cite{Weigel:2013kwa,Weigel:2018xng,Adam:2018tnv}, although recent developments \cite{Manton:2020onl, Manton:2021ipk} show some promising results.
Furthermore, some modifications of the $\phi^4$ model, see e.g.\ \cite{Christ-Lee,Simas:2016hoo,Demirkaya:2017euk},
allow for existence of towers of internal modes localized on the kinks.
Clearly, in such cases a simple resonance energy exchange mechanism
cannot be applied.

Resonance structures can also be observed in the triply-degenerate $\phi^6$
model \cite{Dorey:2011yw}. The kinks in this model do not support localized modes, and the resonance
effects are instead
determined by collective modes 
trapped by the $K\bar K$ system.
A modification of the collective coordinate approach 
reproduces the resonance dynamics of the $\phi^6$ $K\bar K$ collisions qualitatively well \cite{Takyi:2016tnc,Weigel:2013kwa}.

The collective coordinate approximation is limited by the assumption that the modes which
lead to resonances, and the modes of the continuum, are separated.
However, as the frequency of the internal mode of a kink is shifted towards
the mass threshold, an excitation of this mode may also excite the radiative modes of the continuous spectrum,
affecting the resonance exchange mechanism. 
This motivates the current paper, an investigation of 
in $K\bar K$ collisions in a deformation of the sine-Gordon theory, other aspects of which were recently studied in \cite{Dorey:2019uap}.
The deformation lifts the infinite degeneracy of the sine-Gordon vacuum while leaving a $\mathbb{Z}_2$ symmetry
unbroken.
For small values of the deformation parameter, close to sine-Gordon,  
the model possesses two true vacua and a large number of false vacua. These open new channels of 
$K\bar K$ collisions via the excitation of a bubble of false vacuum and its subsequent decay, allowing a smooth transition from the kink-antikink reflection that occurs in the deformed theory to the kink-antikink transmission found in sine-Gordon. At larger values a rich variety of behaviours is found: first, a simple 
deformation of the $\phi^4$ pattern of resonance windows but then, as the kinks of the model acquire more localised modes, a much less regular pattern of windows together with transient structures we call pseudowindows. Finally, close to the critical coupling at which the model has three degenerate vacua, the kinks and antikinks partially fragment into subkinks and their scattering decomposes into a series of pairwise interactions, leading to a novel pattern of windows akin to recent studies of 
the scattering of wobbling kinks in the $\phi^4$ theory \cite{Alonso-Izquierdo:2020ldi}.

The paper is organised as follows.
In Section \ref{sec:Model} we review the deformed sine-Gordon model of \cite{Dorey:2019uap}. We also discuss the dependence of the
spectral structure of linear perturbations of the solutions on the value of the deformation parameter, and the fragmentation of kinks and antikinks which occurs as the $\phi^6$-like limit is approached. 
Section \ref{sec:numerical} explains the numerical methods we used to investigate kink scattering in this model, and gives a birds-eye view of our results.
In Section \ref{sec:Regime1} we present our detailed results for kink-antikink collisions  in the first regime of interest, where the model transitions from sine-Gordon to $\phi^4$ behaviour, and discuss the resonance structures we observed. The collisions of  kinks in the second regime, where the $\phi^4$ model moves to a $\phi^6$-like model, are discussed in Section~\ref{sec:Regime2}. Conclusions and further remarks are contained in the last section.

\section{The Model}\label{sec:Model}
\subsection{Model structure}
We consider the simple theory of real scalar field in 1+1 space-time dimensions specified by the Lagrangian
\be
L=\frac12 \partial^\mu \phi\partial_\mu \phi - U(\phi)
\label{lag}
\ee
where the self-interaction potential $U(\phi)$ is 
\be
U(\phi)= (1-\epsilon)\left(1-\cos \phi \right) +\frac{\epsilon \phi^2}{8\pi^2} \left(\phi-2\pi \right)^2\, ,
\label{pot}
\ee
and $\epsilon $ is a real positive parameter.

\begin{figure}[ht!]
\begin{center}
\setlength{\unitlength}{0.1cm}
\includegraphics[width=.77\textwidth]{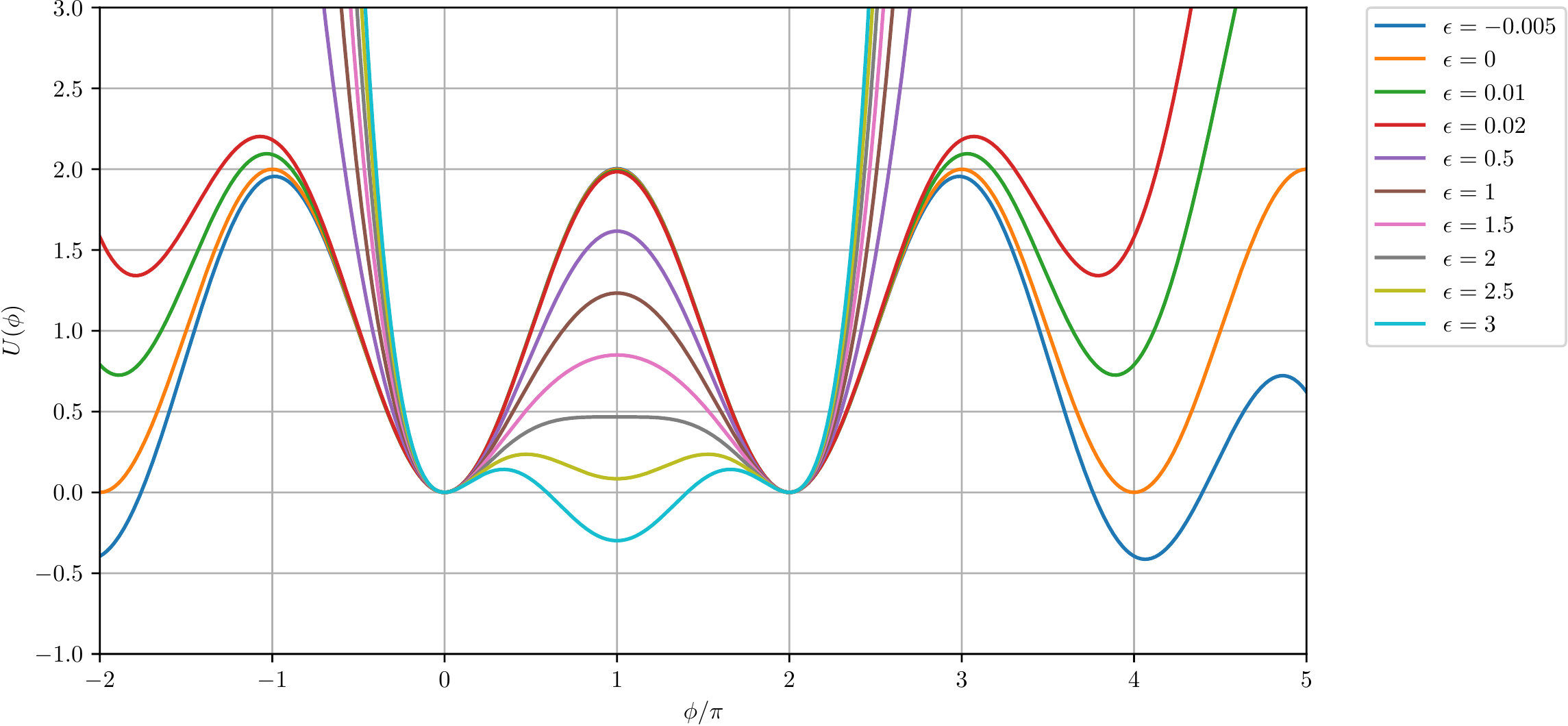}
\end{center}
\vskip -15pt
\caption{The potential of the model \re{lag} for various values of the parameter $\epsilon$. Note that the first and last values plotted are outside the range considered in this paper: for $\epsilon=-0.005$ the potential is unbounded below, while for $\epsilon=3$ it does not support finite-energy kink solutions.}
\label{potentials}
\end{figure}

\begin{figure}[ht!]
\begin{center}
\setlength{\unitlength}{0.1cm}
\includegraphics[width=0.77\textwidth]{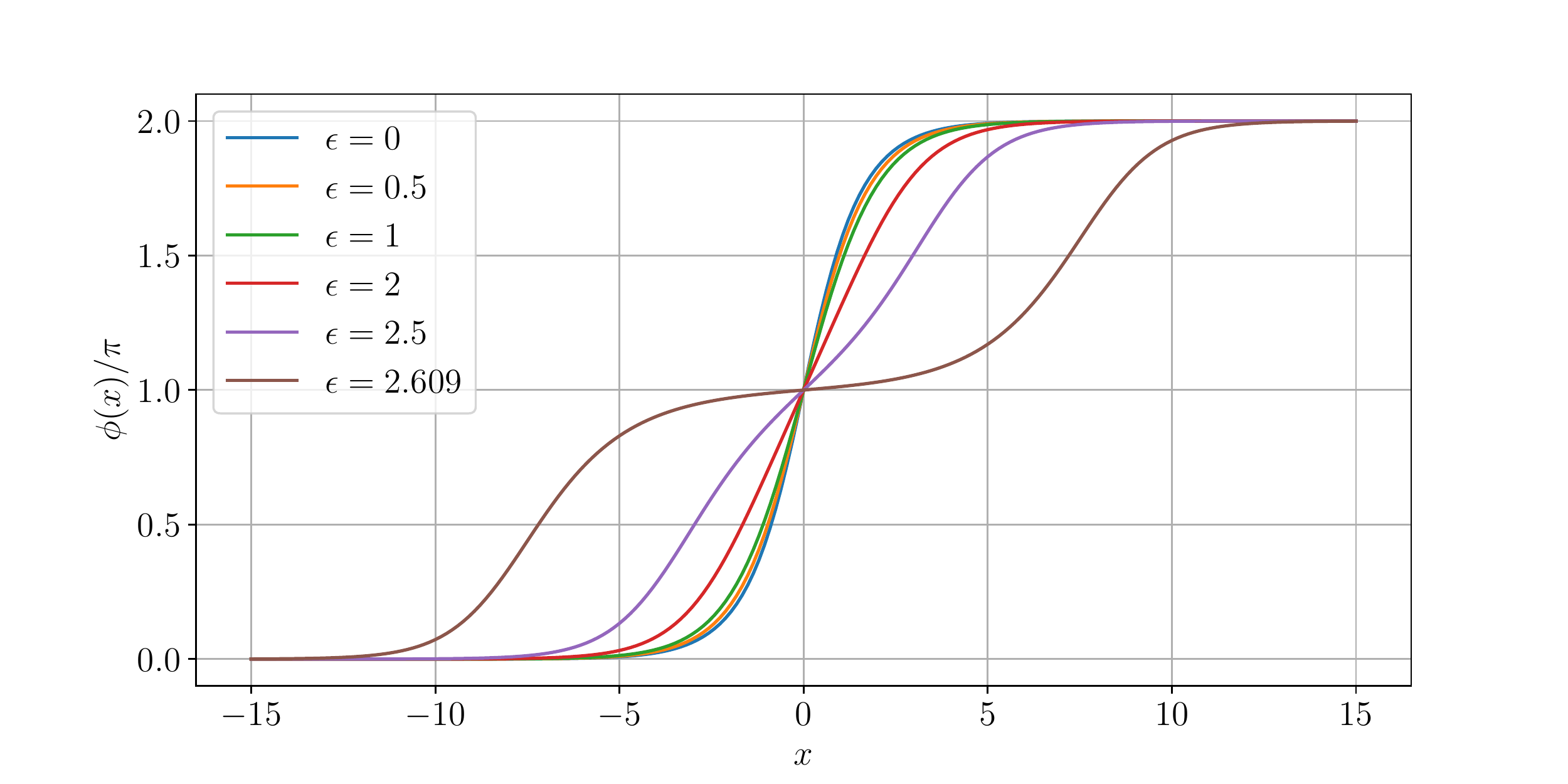}
\end{center}
\vskip -15pt
\caption{Examples of static kink solutions of the model \re{lag} for various values
of the parameter~$\epsilon$. }
\label{profiles}
\end{figure}

This model was proposed in \cite{Dorey:2019uap} to investigate the decay of large amplitude  breather-like states; our goal here is rather
to study the collisional dynamics of its solitons, which turns out to be surprisingly rich.
The parameter $\epsilon$ controls the deformation away from the infinitely-degenerate periodic
sine-Gordon potential, which is recovered in the limit $\epsilon=0$, as shown in figure \ref{potentials}. For any non-zero value of $\epsilon$, the infinite dihedral group
symmetry of the initial potential is broken down to $\mathbb{Z}_2$, generated by a reflection around $\phi=\pi$.
For $\epsilon<0$ the energy is not bounded from below, and we therefore exclude these values. In addition, for $\epsilon>\epsilon_{cr}$, where
\be
\epsilon_{cr}=\frac{16}{16-\pi^2}\approx 2.609945762,
\ee
the potential has a single global minimum at $\phi=\pi$ and the model does not support static kink solutions, so we exclude these values too, and confine our attention to the range $0\le \epsilon \le \epsilon_{cr}$. For all $\epsilon$ in the interior of this range, the potential has just two global minima, at $\phi=0$ and $\phi=2\pi$, and the model is parametrised in such a way that small perturbations around these true vacua always have mass equal to $1$.
In addition to these perturbative states,
the model has  kink and antikink solutions which interpolate between the global minima, the kinks interpolating between $\phi=0$ at $x=-\infty$ and $\phi=2\pi$ at $x=+\infty$ and the antikinks doing the opposite. (We will often be a little loose in our language and refer to topological lumps of either kind as kinks.) Some examples are shown in figure~\ref{profiles}, and our aim in the following is to analyse how they scatter.

In addition to the true vacua, the model's potential has a number of 
the local minima corresponding to false vacua, which can influence kink scattering in important ways.
It is  possible to consider perturbations around these false vacua, as long as they remain small. 
In such cases, for any local minimum $\psi$ of the potential with $U'(\psi)=0$, the square of the mass of small perturbation is $U''(\psi)$. The false vacua vanish when this mass tends to zero. 
These critical points are therefore defined by the pair of nonlinear equations (for $\phi$ and the parameter $\epsilon$)

\begin{equation}
 \frac{\partial U}{\partial \phi}=0,\qquad \frac{\partial^2 U}{\partial \phi^2}=0\,.
\end{equation} 
These critical points allow us to determine the vacuum structure of the model. 
For example for $0.00105685 < \epsilon<0.0074737$  there are four false vacua, near $\phi\approx-4\pi, -2\pi, 4\pi, 6\pi$, while
for $0.0074737<\epsilon<0.0507066$ there are two false vacua (near $\phi\approx-2\pi$ and $4\pi$), and
for $0.0507066<\epsilon<2$ there are only the two true vacua ($\phi=0$ and $\phi=2\pi$).

Setting $\epsilon=1$ yields a shifted and rescaled  potential of the $\phi^4$
model. Increasing  $\epsilon$ further decreases the
height of the potential barrier between the two vacua, and
for $\epsilon>2$ the
local maximum at $\phi=\pi$ splits into two humps located in the intervals $\pi<\phi<2\pi$ and $0<\phi<\pi$, with a
local minimum between them at $\phi=\pi$. For $2<\epsilon<\epsilon_{cr}$ this minimum corresponds to a false vacuum, and for
$\epsilon$ in this range the potential \re{pot} has a very similar shape to that of the
$\phi^6$  model considered many years ago by Christ and Lee~\cite{Christ-Lee}. The mass of small fluctuations about this false vacuum is
\begin{equation}
    m_1(\epsilon)=\sqrt{U''(\pi)}=\sqrt{(\epsilon-2)/2}\,.
\end{equation}
Finally, at $\epsilon=\epsilon_{cr}$ the vacuum at $\phi=\pi$ becomes degenerate with those 
at $\phi = 0$ and $2\pi$, and the model resembles the triply-degenerate $\phi^6$ model that was found to exhibit resonant scattering in \cite{Dorey:2011yw}.

The field equation of the model \re{lag} can be written as
\be
\phi_{tt} - \phi_{xx} + (1-\epsilon)\sin\phi +\frac{\epsilon \phi}{2 \pi^2}\left(\phi-\pi \right)\left(\phi- 2 \pi \right) = 0\,.
\label{eqs}
\ee

For all $\epsilon\in [0,\epsilon_{cr})$, the static kink $\phi_K(x;\epsilon)$ depends parametrically  on $\epsilon$ and
interpolates between the vacua $\phi=0$ as $x\to -\infty$ and $\phi=2\pi$ as $x\to \infty$. In two cases
its  form is known analytically:
\be
\phi_K(x;\epsilon=0)= 4 \arctan e^{x-x_0}\,;\qquad \phi_K(x;\epsilon=1)= \pi \left( \tanh\frac {x-x_0}{2} + 1 \right).
\ee
There is little visual difference between these two solutions, as can be seen from figure \ref{profiles}.
Despite their similarity, the dynamical properties of these kinks are very distinct, as will be seen below. 
Stronger deformations of the potential \re{pot} significantly affect not just the scattering but also 
the form of the topological solitons, and as
the deformation parameter approaches the critical value $\epsilon_{cr}$, they become almost decomposed into
pairs of subkinks interpolating between the true vacua at $\phi=0$ and $2\pi$ via the false vacuum at $\phi=\pi$,
as also shown in figure \ref{profiles}. In the limit $\epsilon\to\epsilon_{cr}$ the potential becomes triply degenerate and the kinks split into two, in the corresponding different topological sectors.

\begin{figure}[ht!]
\begin{center}
\includegraphics[width=0.77\textwidth]{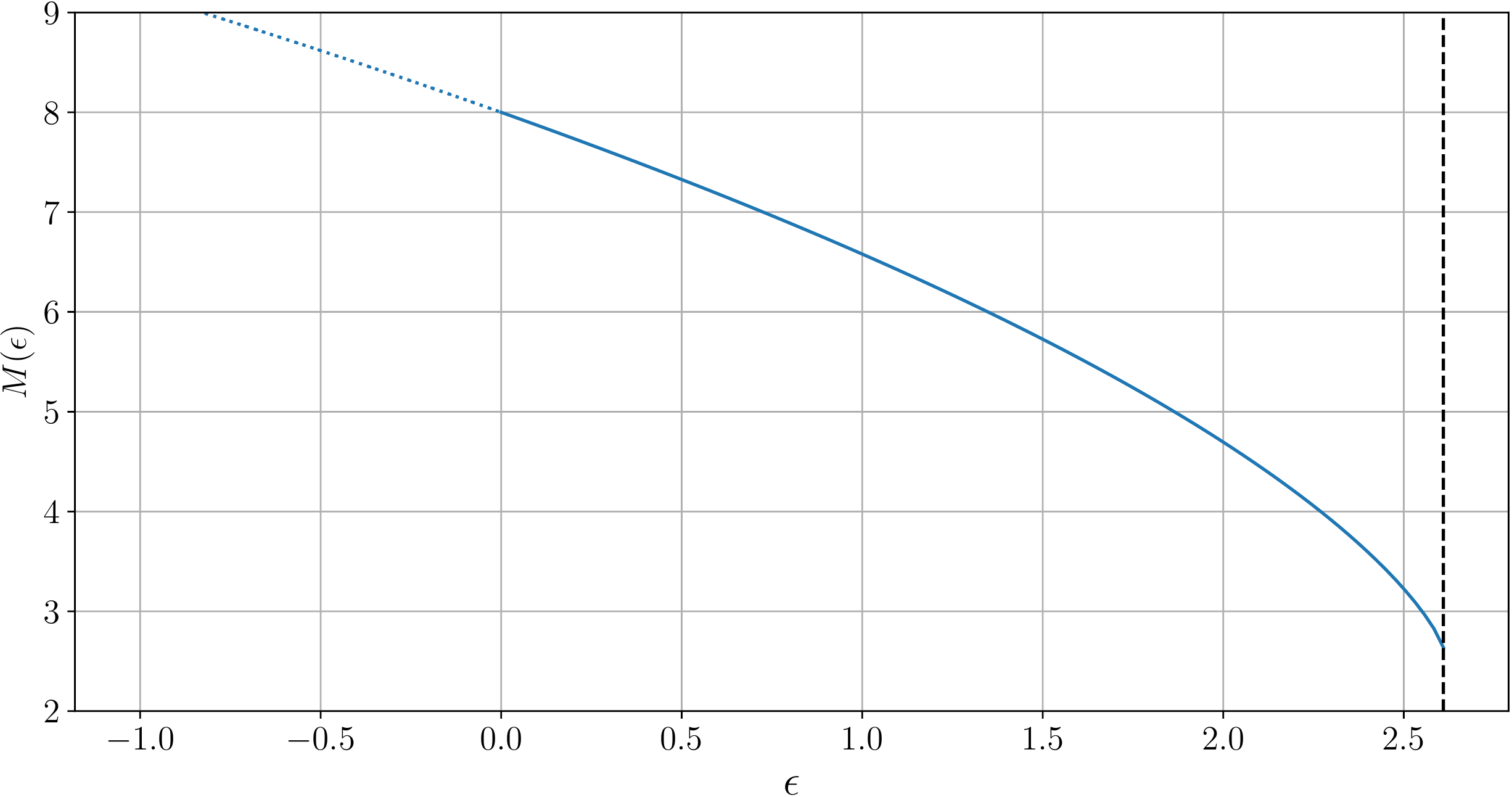}
\end{center}
\vskip -10pt
\caption{The mass $M(\epsilon)$ of a static kink as a function of the deformation parameter $\epsilon$.}
\label{Energy}
\end{figure}

The mass of the kink,
\begin{equation}
    M(\epsilon)=\int_{0}^{2\pi}\sqrt{2U(\phi)}\,d\phi\,,
\end{equation}
can be obtained from the standard Bogomolnyi' trick. It
decreases from $M(0)=8$ through $M(1)=\frac{2\pi^2}{3}\approx 6.5797$ to $M(\epsilon_{cr})\approx 2.6363$ (cf.\ figure \ref{Energy}). Note that this limiting mass as $\epsilon\to\epsilon_{cr}$ 
is exactly twice the mass of a `true' single kink of the model at $\epsilon=\epsilon_{cr}$. 
The Bogomol'nyi
trick also reduces the second order static field equation to the first order Bogomol'nyi–Prasad–Sommerfield (BPS) equation
\begin{equation}
\phi_x=\sqrt{2U(\phi)}
\label{BPSeq}
\end{equation}
which is much easier to solve and requires only a single integration constant. The profiles assume the vacuum values at spatial infinities.

\subsection{Spectral structure}
\label{SpectStruct}

\begin{figure}[ht!]
\begin{center}
\setlength{\unitlength}{0.1cm}
\includegraphics[width=0.8\textwidth]{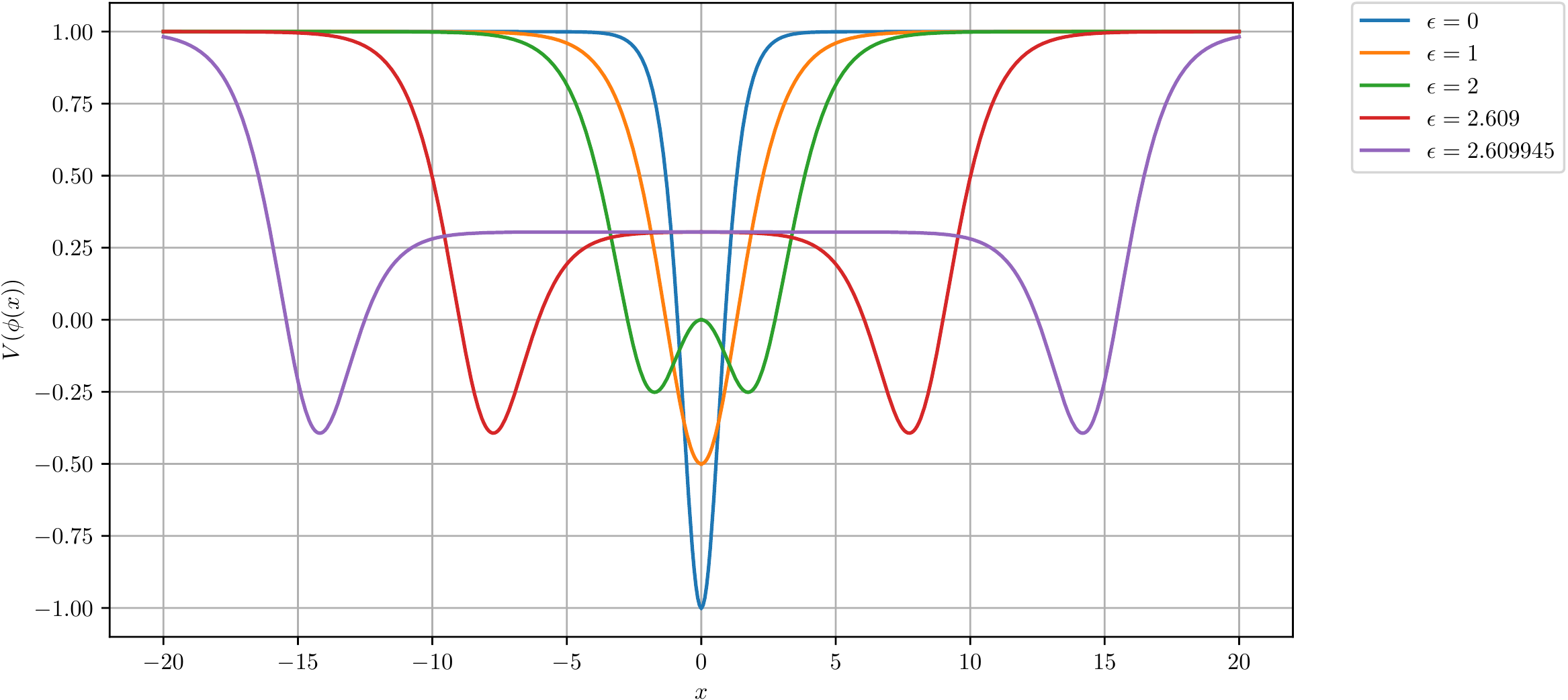}
\end{center}
\vskip -10pt
\caption{The linearized potential $V(x)$ \re{eff-pot} for various values of the deformation parameter $\epsilon$.}
\label{linear_potentials}
\end{figure}

We now consider the spectrum of linear perturbations of a single kink by setting $\phi(x,t)=\phi_K(x;\epsilon) + \eta(x,t)$
where $\eta(x,t)=\xi(x)e^{i\omega t}$. The corresponding equation for the
fluctuation eigenfunctions is
\be
\left( -\frac{d^2}{dx^2 } +
V(x)
\right)\xi(x) = \omega^2\xi(x)
\label{pert}
\ee
where
\be
V(x)=U''(\phi_K(x))
=(1-\epsilon) \cos \phi_K(x)
+  \frac{\epsilon}{2\pi^2}\left( 3(\phi_K(x)-\pi)^2-\pi^2\right)
\label{eff-pot}
\ee
is the corresponding effective potential for the linear excitations, displayed in figure \ref{linear_potentials} for various representative values of $\epsilon$.
Note that the parameters
of the model \re{lag} are fixed in such a way that the masses of the
excitations around the true vacua $\phi=0$ and $\phi=2\pi$
remain the same and equal to $1$, as in the original sine-Gordon model.

\begin{figure}[ht!]
\begin{center}
\includegraphics[width=0.77\textwidth, angle =0]{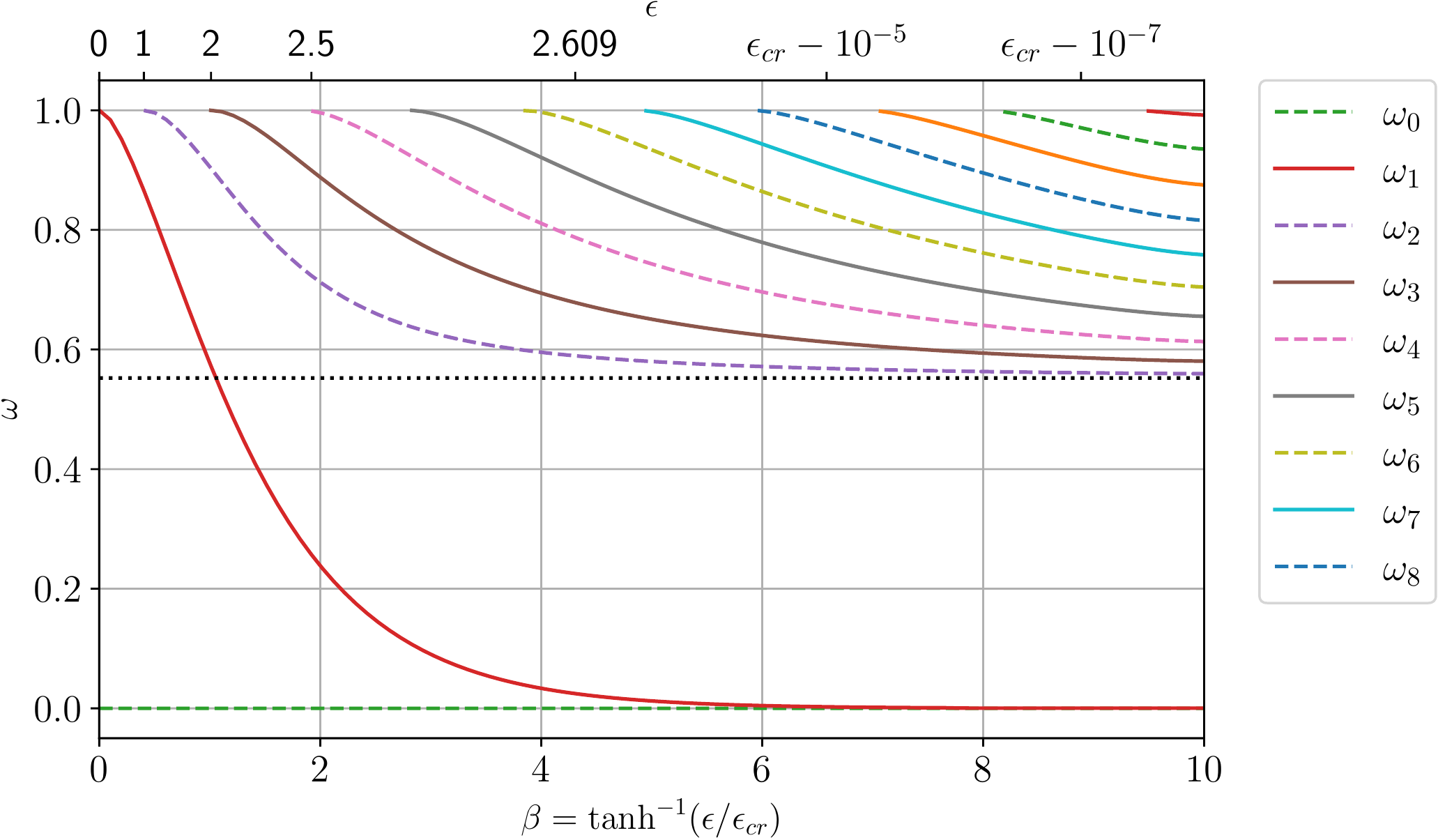}
\end{center}
\vskip -10pt
\caption{The eigenfrequency of the internal modes
$\omega$ as a function of the deformation parameter $\epsilon=\epsilon_{cr}\tanh(\beta)$. Dashed lines correspond to even modes. The dotted line shows $m_1(\epsilon_{cr})$, the mass associated with small perturbations of the false vacuum at $\epsilon=\epsilon_{cr}$. }
\label{spectral_structure}
\end{figure}

We can solve (\ref{pert}) numerically for each value of the deformation parameter $\epsilon$.
For $\epsilon=0$ the sine-Gordon model is undeformed, and the spectrum of linearized kink
excitations does not contain any
internal modes, but rather just the usual translational zero mode and the states of the continuum.
These modes are not
affected by the variation of the parameter $\epsilon$, and exist for all of its values.
However, as $\epsilon$  increases from zero an internal mode appears,
its eigenfrequency  $\omega_1$ smoothly decreasing from the mass threshold with increasing $\epsilon$ as shown in
figure \ref{spectral_structure}, where we used the transformed parameter \be\beta=\tanh^{-1}(\epsilon/\epsilon_{cr})
\ee
for better visibility of the spectrum near the critical value $\epsilon=\epsilon_{cr}$. 

For $\epsilon=1$ the effective potential \re{eff-pot} reduces to the well-known P{\"o}schl-Teller
potential for the $\phi^4$ theory. In this case the corresponding spectrum of fluctuations
contains the translational zero mode $\omega_0=0$, one internal mode of the kink with
frequency $\omega_1 = \sqrt 3/2$, and
the continuum
modes $\omega_k=\sqrt{k^2+1}$.

For $\epsilon>1$ more internal modes appear, their number tending to infinity as $\epsilon$
approaches $\epsilon_{cr}$.
Recall that for $2<\epsilon<\epsilon_{cr}$ there is a false vacuum at $\phi=\pi$.
As can be seen in figure~\ref{profiles} and as analysed in more detail in the next subsection, kinks in this region come to look like a pair of bound subkinks, the first interpolating $0\to\pi$ and the second
$\pi\to2\pi$, rather than a single kink $0\to 2\pi$.
Each of these subkinks generates its own potential well, but these do not support localised modes. However
the presence of the false vacuum yields an almost
flat potential between the subkinks, as shown in figure \ref{linear_potentials}. As $\epsilon\to\epsilon_{cr}$ the distance
between the two subkinks tends to infinity.
This limit corresponds to a very wide potential well, within which many bound modes can be trapped.
At $\epsilon=\epsilon_{cr}$, the false vacuum $\phi=\pi$ becomes a true
vacuum and the subkinks become independent topological defects.
These new, smaller kinks are not symmetric: they interpolate between one of the original two vacua $\phi=0$ or
$\phi=2\pi$, for which the mass of the linearized perturbations is $m=1$,
and the new third vacuum at $\phi=\pi$. The mass of excitations about this
vacuum is different, \mbox{$m_1(\epsilon_{cr})=\sqrt{\frac{\pi^2-8}{16-\pi^2}}\approx0.5522<1$}.
As can be seen by comparing figure \ref{halfkinks} below with the corresponding plots in \cite{Dorey:2011yw}, the collisional 
dynamics of these new kinks is close to the resonant $K\bar K$ scattering
observed in the $\phi^6$ model, where the resonance
structure appears not because of the presence of internal modes of the kinks but
rather from collective modes trapped between them.

\subsection{The double kink}
A kink can fragment into subkinks when one (or more) false vacua lie between two true vacua. This occurs in our model
as $\epsilon\to\epsilon_{cr}$, the single kink fragmenting into two subkinks separated by a region of false vacuum. The field of the subkinks situated at $x=\pm L$ approaches the false vacuum $\psi(\epsilon)$ as
\begin{equation}
    \phi(x) - \psi(\epsilon) \approx \pm \frac{B}{m_1(\epsilon)}e^{\pm m_1(\epsilon)(x\mp L)}.
\end{equation}
The constant $B$ can be obtained from the numerical solution of the BPS equation (\ref{BPSeq}). 
Its value depends on how we choose to define the positions of the subkinks. One natural choice is the point at which the field reaches the local maximum of potential. For $\epsilon=\epsilon_{cr}$ this is at $\phi=1.367646$ and $\phi=2\pi-1.367646=4.915539$, for which $B=1.210404$.

The force acting on the subkinks is a sum of the standard kink-kink repulsion (see for example \cite{Manton:2004tk}) and the negative pressure of the false vacuum:
\begin{equation}
    F = 2B^2e^{-2m_1(\epsilon)L} - U(\psi(\epsilon)).
\end{equation}
In this case $\psi(\epsilon)=\pi$ for all $\epsilon\in (2,\epsilon_{cr})$ and
\begin{equation}\label{critical_U}
    U(\psi(\epsilon))=
    \frac{16-\pi^2}{8}(\epsilon_{cr}-\epsilon)
\end{equation}

Static solutions balance between the scalar repulsion and false vacuum pressure, and the condition $F=0$ gives the separation between the subkinks to be
\begin{equation}\label{double_kink_size}
    L(\epsilon) = -\frac{1}{2m_1(\epsilon)}\log\left[\frac{U(\psi(\epsilon))}{2B^2}\right]\,.
\end{equation}
Keeping only the lowest terms in $\epsilon_{cr}-\epsilon$ and using
\mbox{$m_1(\epsilon_{cr})=\sqrt{\frac{\pi^2-8}{16-\pi^2}}$} 
we obtain

\begin{equation}
    L(\epsilon) = -\frac12\sqrt{\frac{16-\pi^2}{\pi^2-8}}\log\left[\frac{16-\pi^2}{16B^2}(\epsilon_{cr}-\epsilon)\right]
\end{equation}
or, in terms of $\beta$ where near the critical value of $\epsilon$ we can use $\epsilon_{cr}-\epsilon\approx 2e^{-2\beta}$,

\begin{equation}
    L(\beta) = \frac12\sqrt{\frac{16-\pi^2}{\pi^2-8}}\left[2\beta-\log\frac{16-\pi^2}{8B^2}\right].
\end{equation}
This formula approximates the positions of the subkinks very well for large values of $\beta$.

\subsection{Small perturbations of the double kink}\label{sec:frequencies}\label{doublekinkoscillation}

The static solution representing a double kink balances the force repelling the subkinks and the attractive force due to the false vacuum trapped between them. 
When the subkinks are moved away from their equilibrium positions, say by some value $\delta L$, they will start to oscillate around these positions.  Small displacements result in a returning force

\begin{equation}
    \delta F=\frac{d F}{dL}\delta L =  2B^2e^{-2m_1L}(-2m_1)\delta L
\end{equation}
but since at equilibrium $F=0$ we can get rid of the constant $B$  and write
\begin{equation}
    \delta F = -2m_1U(\psi)\delta L.
\end{equation}
By taking into account that the mass of the subkink is a half of the mass of the full kink $M_{1/2}=M(\epsilon)/2$ we can find the frequency of the small oscillations:

\begin{equation}\label{frequencies}
\omega^2=\frac{4m_1U(\psi)}{M(\epsilon)}.
\end{equation}
The leading term as $\epsilon\to\epsilon_{cr}$ is
\begin{equation}
    \omega=\sqrt{\frac{(16-\pi^2)(\epsilon_{cr}-\epsilon)}{2M(\epsilon_{cr})}}\approx0.801\sqrt{\epsilon_{cr}-\epsilon}\approx1.83\,e^{-\beta}.
\end{equation}

These oscillations move the subkinks closer together and then further apart, corresponding to an antisymmetric excitation of the translational modes of the subkinks.  This motion corresponds to the excitation of the first non-zero mode $\omega_1$ of the full double kink, shown as a solid red curve on figure \ref{spectral_structure}. We have checked numerically within the range $\epsilon_{cr}-\epsilon\in[10^{-5},10^{-2}]$, and $\omega_1(\epsilon)$ indeed scales as the square root of the difference $\epsilon_{cr}-\epsilon$ in this region, its numerical value agreeing with the approximation to at least three significant figures.

\subsection{Further static solutions: the unstable lump}\label{sec:unstable_lumps}
The presence of a false vacuum 
allows for another class of static solutions, albeit with infinite energy. These
involve subkinks and subantikinks which interpolate between neighbouring true and false vacua.
In scalar field theories a kink and an antikink always attract each other. However, if they are surrounded by a false vacuum with a true vacuum trapped in between, there is an additional force trying to separate them and thereby expand the region of true vacuum. For wide separations this force is constant, depending only on the difference between the field theoretic potentials in the false and true vacua. Since the attractive kink - antikink force decays exponentially,
this opens a possibility that there is a single distance where the two forces  balance. Note that perturbation in any direction would lead to growing acceleration, destroying the unstable static state.

In contrast to the situation for the double kinks, neighbouring true and false vacua are found both for $\epsilon$ small and for $\epsilon\to\epsilon_{cr}$.
Suppose the relevant false vacuum is at $\phi=\psi$. In order to find the unstable solution we multiply the static version of 
(\ref{eqs}) by $\phi_x$ and integrate once, imposing the false vacuum boundary condition $\phi=\psi$ at spatial infinity to find the BPS-like
equation
\begin{equation}\label{BPS2}
    \phi_x=\pm \sqrt{2[U(\phi)-U(\psi)]}\,.
\end{equation}

\begin{figure}[ht!]
\begin{center}
\includegraphics[width=0.49\textwidth]{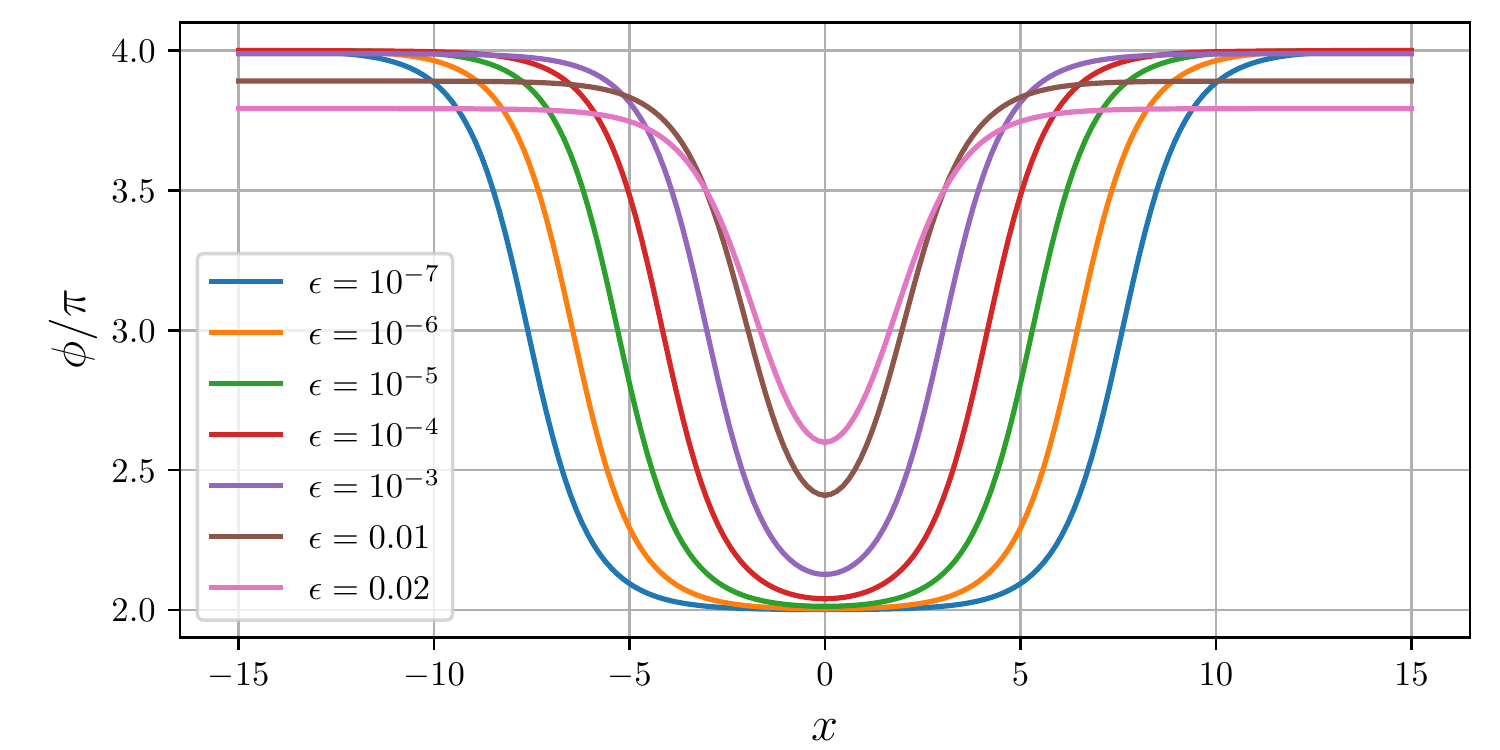}
\includegraphics[width=0.49\textwidth]{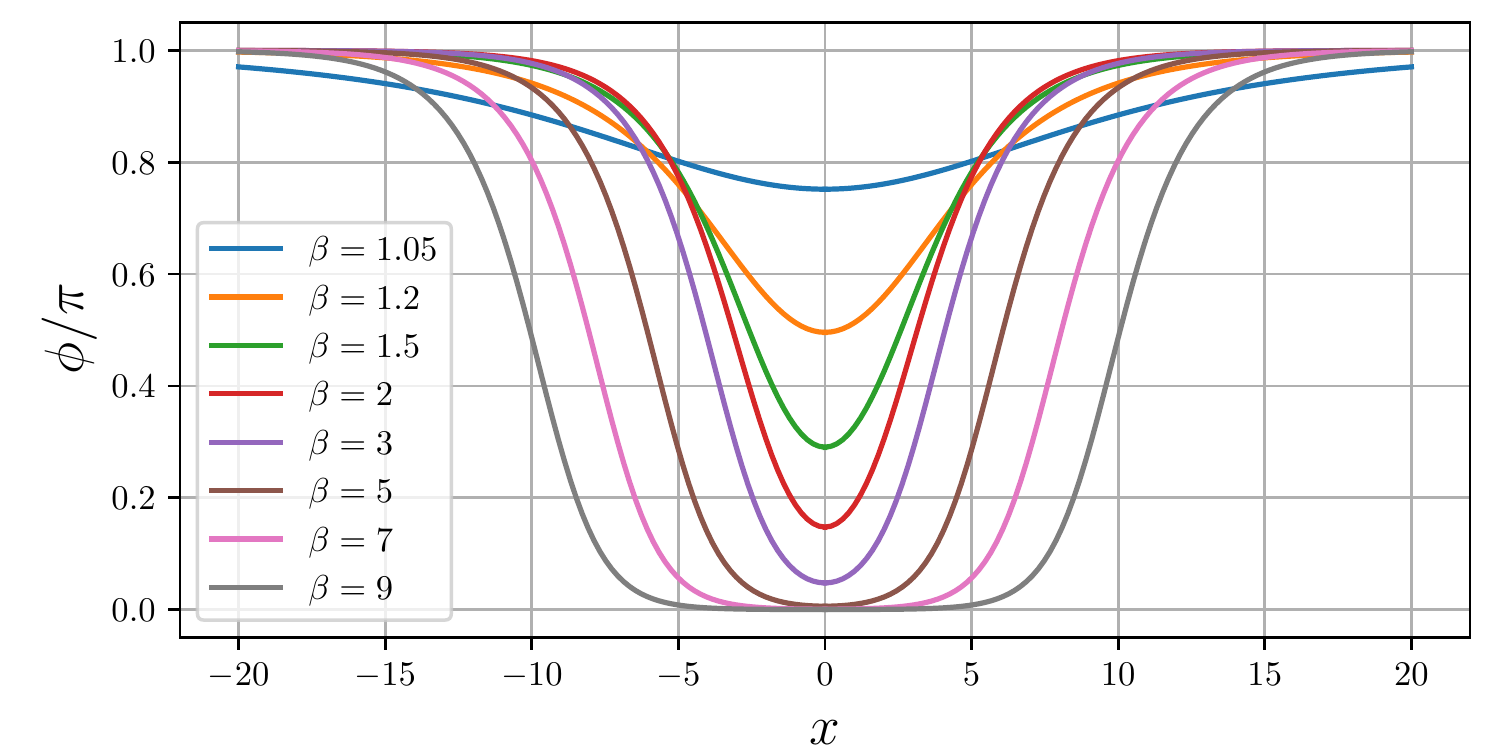}
\end{center}
\caption{Profiles of unstable lumps. \textit{Left:} unstable lumps for small values of $\epsilon$ based on the first false vacuum $\psi>2\pi$; \textit{right:} unstable lumps for $\epsilon\to\epsilon_{cr}$ based on the false vacuum at $\psi=\pi$ and parametrised by $\beta$, with $\epsilon=\epsilon_{cr}\tanh(\beta)$.
.}
\label{unstable_profiles}
\end{figure}

The full-line solutions to this equation have zero topological charge and cannot be kinks: rather, they are unstable lumps, which can be viewed as static
(sub-) kink and antikink pairs, as described in the opening paragraph of this section. The derivative $\phi_x$ vanishes before reaching the true potential minimum at one of the true vacua. Therefore we can apply the boundary condition at this point, thereby defining the position of the unstable solution:
\begin{equation}
U(\phi(0))=U(\psi)\,.
\end{equation}
This is a transcendental equation for $\phi(0)$ and has to be solved numerically. But having the value $\phi(0)$ we can integrate the equation (\ref{BPS2}) with different signs for $x>0$ and $x<0$. Example solutions are shown in figure \ref{unstable_profiles}. Note that both for small (left panel) and near-critical (right panel) values of $\epsilon$, the unstable solution develops an antikink-kink structure as the potential difference between the false and true vacua decreases.

For large separations (small values of $\epsilon$ or $\epsilon_{cr}-\epsilon$) we can use a similar approximation to that for the double kink and find the positions of the kink and antikinks with a few substitutions.
The false vacuum is now on the outside pulling the kink and antikink apart and the  true vacuum is trapped in between. Therefore in equation (\ref{double_kink_size}) $m_1$ has to be replaced by the mass of the true vacuum $m=1$. For small values of $\epsilon$ the vacuum at $\phi=4\pi$ is raised so that $U(4\pi)\approx 4\pi^2\epsilon$. Since the profiles of the kinks are known in the unperturbed model we can calculate $B=4$ explicitly. This leads to 
\begin{equation}
    L(\epsilon) = -\frac12\log\left(\frac{\pi^2\epsilon}{4}\right)
\end{equation}
which agrees well with the numerically found profiles. 
In the near critical case $U(\psi(\epsilon))$ is also known (\ref{critical_U}), but $B=3.003373$ had to be determined numerically

\begin{equation}
    L(\epsilon) = -\frac12\log\left[\frac{16-\pi^2}{16B^2}(\epsilon_{cr}-\epsilon)\right]= \frac12\left[2\beta-\log\frac{16-\pi^2}{8B^2}\right].
\end{equation}

In the case of unstable lumps the forces act in opposite directions than for double kinks, and equation (\ref{frequencies}) for the frequencies should therefore have the opposite sign (the true vacuum is inside and the false one
outside), and instead of $m_1$ we can put $1$ to yield
\begin{equation}
    \omega^2=-\frac{4U(\psi)}{M(\epsilon)}.
\end{equation}
For $\epsilon\to 0$ we can use the mass of the sine-Gordon kink $M=8$ and the limiting eigenfrequency of the unstable mode is therefore $\omega^2=-4\pi^2\epsilon$.
For $\epsilon\to\epsilon_{cr}$ we rather obtain
\begin{equation}
    \omega^2=-\frac{16-\pi^2}{M(\epsilon_{cr})}(\epsilon_{cr}-\epsilon)
    \approx -2.325(\epsilon_{cr}-\epsilon) \approx -12.14\,e^{-2\beta}\,.
\end{equation}

On their own, these unstable lumps have infinite energy, because of the false vacuum surrounding them. However, as will be seen below, at least for $\epsilon\to\epsilon_{cr}$ they can still play an important role in the dynamics of the system when considered locally. Note that for $\epsilon\to 0$ there are further unstable lump solutions to the static field equations corresponding to a region of false vacuum embedded in a region of even more false (higher energy) vacuum. These can be found and analysed exactly as above, but as we did not yet find a role for them in the scattering processes under consideration in this paper we will not discuss them further here.

\section{Numerical methods and overall results}\label{sec:numerical}
We used the method of lines to solve the second order evolution equation (\ref{eqs}), discretizing the spatial part using the standard five point stencil on a uniform grid. Given the even parity of our initial conditions we  solved on a half interval, implementing the appropriate boundary condition $\phi(-x)=\phi(x)$ at $x=0$. For the second boundary condition we used the standard reflecting boundary, making sure to stop the evolution before any radiation could be reflected from the boundary and return to the centre of collision to interfere with the results. (Placing the second boundary at $x=T_s/1.8$, with $T_s$ the total run-time of the simulation, was typically enough for this.)
For the time stepping function we used a fourth order symplectic method which conserves energy and is reliable even for long time evolution for isolated hamiltonian systems.
The spatial distance between the grid points was usually $dx=0.05$ but we checked smaller values to verify the consistency of the results. The time step was usually set as $0.4dx$, but again, we also checked other values to confirm the convergence of our results.

\begin{figure}[ht!]
\begin{center}
\includegraphics[width=1.05\textwidth]{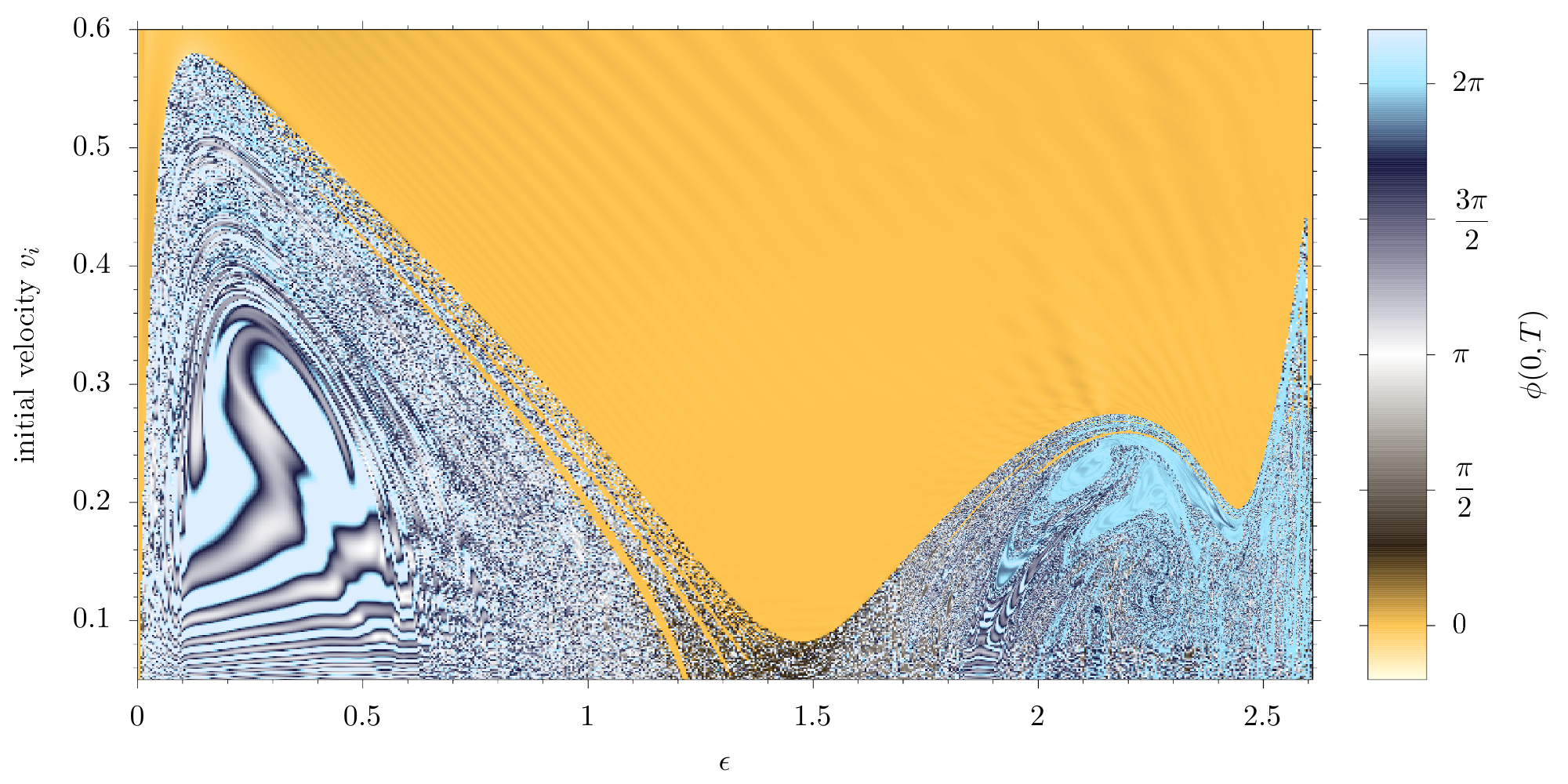}
\end{center}
\vskip -20pt
\caption{The post-collision value of the field at the collision centre
as a function of the initial velocity of the colliding kinks and the
deformation parameter $\epsilon$. The field value was measured $300$ units of time after the intersection of the initial trajectories, that is $300+x_0/v_i$ units of time after the start of the simulation. Figures
\ref{nearcritscan} and
\ref{nearcritscan_zoom} below zoom in on the right-hand edge of this plot.}
\label{fullscan}
\end{figure}

Figure \ref{fullscan} displays our results for 
$\bar K K$ collisions with initial conditions taken to be an antikink at $x=-x_0=-20$ and a kink at $x=x_0=20$, with  velocities $v_i$ and $-v_i$ respectively, for the
full range of values of $\epsilon \in [0,\epsilon_{cr}]$. The figure shows the post-collision value of the field at the centre of collision, as a function of $v_i$ and the perturbation parameter $\epsilon$. The field is measured $300$ units of time after the moment of intersection of the initial antikink and kink
trajectories, that is at $300+x_0/v_i$ units of time from the start of the simulation. 

Since the spectrum of linearized perturbations about a single kink depends strongly
on the value of the deformation parameter
$\epsilon$, it is no surprise that the collisional dynamics of the kinks revealed in the figure has an intricate and varied structure.
This structure is particularly rich near $\epsilon=\epsilon_{cr}$, and zoomed-in plots of this region are shown in figures
\ref{nearcritscan} and
\ref{nearcritscan_zoom} below. (For the last of these, figure \ref{nearcritscan_zoom}, the increased widths of the antikink and kink in the regime plotted made it necessary to set $x_0=40$ to ensure that they were initially well separated.)

Figure \ref{fullscan} naturally splits into two parts: regime 1, $0\leq\epsilon \lesssim 1.45$, which includes the interpolation between $\phi^4$ and sine-Gordon like scattering, and regime 2, $1.45\lesssim \epsilon\leq \epsilon_{cr}$, where the false vacuum at $\phi=\pi$ enters the picture and comes to play an important r\^ole as $\epsilon$ approaches $\epsilon_{cr}$.
In the following two sections we will analyse these regimes in turn.

\section{The first transition: from sine-Gordon to $\phi^4$ scattering}\label{sec:Regime1}

\subsection{False vacuum effects for small $\epsilon$}
At $\epsilon=0$ the model is integrable and scattering is perfectly elastic, and the vacua at $0$ and $2\pi$ are exactly degenerate with further vacua at $-2\pi$, $4\pi$ and so on. An initial kink - antikink pair interpolating from $0$ to $2\pi$ and then back to $0$ scatters to an oppositely-ordered antikink - kink pair for which the intermediate vacuum is $-2\pi$. For any nonzero value of $\epsilon>0$ such a final state is impossible, as value of the potential \re{pot} at $\phi=-2\pi$ no longer degenerate with that at $\phi=0$. Nevertheless, 
for small values of the parameter $\epsilon \lesssim 0.05$
the potential retains a false vacuum
at $\phi\approx-2\pi$ (along with its $\mathbb{Z}_2$ reflection at $\phi\approx 4\pi$), as illustrated in figure~\ref{potentials}.
The presence of such false vacua significantly affects the dynamics of the
kinks \cite{Ashcroft:2016tgj,Gomes:2018heu}. In the model under consideration
the collision channels associated with them enable a smooth transition from the situation for $\epsilon> 0$ back to
sine-Gordon scattering at $\epsilon=0$.

\begin{figure}[!ht]
 \begin{center}
\setlength{\unitlength}{0.1cm}
\includegraphics[width=\textwidth]{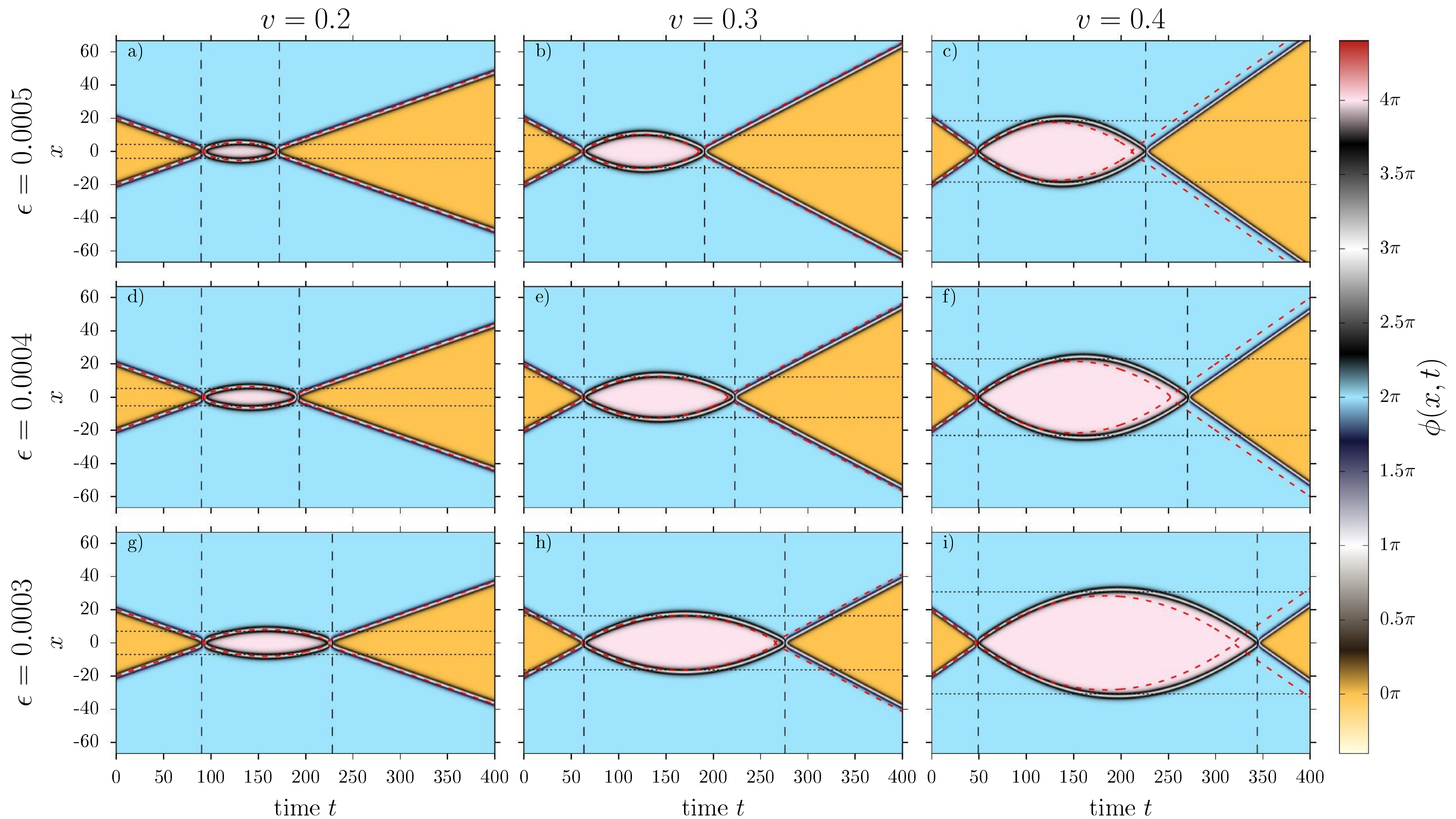}
\end{center}
\caption{Collisions between an antikink and a kink with excitation of the false vacuum
in the deformed model \re{lag}.
The plots represent the field values as functions of the impact velocity and the time. Black dashed lines show the theoretically predicted bubble size and life-time. Red dashed lines show the paths calculated from the effective model (\ref{effective}). Note that the effective model is non-relativistic, and this is reflected in the worsening of its predictions with increasing velocities.}\label{false_collisions}
\label{fig9}
\end{figure}

Figure \ref{fig9} shows a collection of
space-time maps of
the scalar field through the collision process.
As the impact velocity becomes
large enough to overcome the potential barrier, a quasi-elastic sine-Gordon like scattering of the kinks with a flipping
of the central vacuum from true to false is observed. In other words, the radiative losses are almost negligible, the collision of the solitons produces a bubble of false vacuum, with
the outgoing kinks bounding the false vacuum domain within the true vacuum. For small values of
$\epsilon$ the difference between the true and false vacuum energies is small and the dynamics of the outgoing kinks can be treated using a thin wall approximation as in the problem of the decay of false vacuum, see
\cite{Kobzarev:1974cp,Coleman:1977py,Voloshin:1985id}. 

After the initial collision the new kinks bounding the false vacuum region move apart and decelerate 
\cite{Kiselev:1998gg,Gomes:2018heu}, stopping at the turning points $x=\pm L/2$ where
the volume energy $16\pi^2\epsilon L$ of the bubble together with the bubble wall surface energy $E_\sigma$
becomes equal to the initial kinetic energy of the colliding kinks, i.e.
\be
E_\sigma + 16\pi^2\epsilon L = \frac{2 M(\epsilon)}{\sqrt{1-v_{in}^2}}\,.
\ee
Taking  into account that in the thin wall approximation the mass of the kink is not very different from the mass of the
sG kink $M=8$, and $E_\sigma\approx 2 M$, we find the maximal separation between the kinks bounding the false vacuum is
\be
L=\frac{1}{\pi^2\epsilon}\left(\frac{1}{\sqrt{1-v_{in}^2}}-1\right).
\ee
This is consistent with deceleration from the velocity $v$ to $0$ with a constant rate of $a_0=\pi^2\epsilon$.
From \cite{Kiselev:1998gg} and \cite{Manton:2004tk} we know that topological defects experience a force equal to the difference between the field potential levels of the true and false vacua. In our case 
\begin{equation}
    F=U(4\pi)-U(2\pi)=U(0)-U(-2\pi)=8\pi^2\epsilon=Ma_0\,.
\end{equation}
The lifetime of the false vacuum bubble can also be calculated using the relativistic deceleration, giving
\begin{equation}
    t_{bubble}=2\sqrt{\frac{L}{a_0}(2+a_0L)} = \frac{2v}{\pi^2\epsilon\sqrt{1-v^2}}\,.
\end{equation}

The kinks then collide for a second time, and the resurrected $K \bar K$ pair finally separates in the
true vacuum. The radiative energy loss for $\epsilon \ll 1$ is small, and thus the final escape velocity
$v_{out} \approx v_{in}$. In the sine-Gordon limit $\epsilon \to 0$ the vacua become degenerate, $t_{bubble}$ diverges, and
the kinks collide perfectly
elastically with a flip of the vacuum $2\pi \to - 2\pi$ in the final state so that sine-Gordon scattering is restored.

\subsection{Effective model}
Figure \ref{false_collisions} shows that as $\epsilon\to 0$ at fixed $v$ the individual collisions become almost elastic, so we can neglect all radiation effects as well as the coupling to the bound mode. 
On the other hand, the false vacuum continues to be important: as just described,
for all non-zero $\epsilon$ it leads to a recollision a time $t_{bubble}$ after the initial collision, returning the vacuum at the origin to its original value of $2\pi$, while for $\epsilon=0$ there is no second collision, and the vacuum at the origin retains its flipped value of $-2\pi$ in the final state. 
It is straightforward to construct and study an effective model in this limit of $\epsilon$ small but nonzero.  For $\epsilon=0$, analytic solutions describing $K\bar K$ collisions are known. For non-relativistic initial velocities we can adopt a moduli space approximation \cite{Manton:2020onl}
\begin{equation}\label{moduli}
    \phi(x,t)=4\arctan\left(\frac{\sinh a(t)}{\cosh x}\right).
\end{equation}
For $a$ and $x$ large and positive, $\phi(x,t)\approx 4\arctan\left(e^{a-x}\right)$ and approximates a single antikink located at $x=a$; similar considerations in the other quadrants confirm that 
 $a$ is a modulus which asymptotically corresponds to the position of the antikink in the solution, with $-a$ being the position of the kink. In the sine-Gordon model, nonrelativistic scattering with velocity $v\ll 1$ is reproduced by setting $a(t)=a_{sg}(t)$ with
\begin{equation}
    \sinh a_{sg}(t)=\frac{\sinh vt}{v}.
\end{equation}
For small values of $\epsilon$ we can assume that the solution (\ref{moduli}) approximates well the instantaneous shape of the field in the deformed model, replacing the above formula for $a_{sg}(t)$ with a general function $a(t)$. 
Putting this into the Lagrangian density and integrating over $x$ we obtain a Lagrangian describing a single particle 
\begin{equation}\label{effective}
    L=\frac12 g(a)\dot a^2 - (1-\epsilon) W_{sG} - \epsilon W_{4}
\end{equation}
where
\begin{equation}
    g(a) = \int_{-\infty}^{\infty}\left(\frac{\partial\phi}{\partial a}\right)^2dx=16\left(1+\frac{2a}{\sinh(2a)}\right)
\end{equation}
is a metric on the moduli space which plays the role of an effective mass, and
\begin{equation}
W_{sG,4} = \int_{-\infty}^{\infty}\left[\frac12\left(\frac{\partial\phi}{\partial x}\right)^2+U_{sG, 4}(\phi)\right] dx
\end{equation}
is an effective potential where $U_{sG}(\phi) = 1-\cos\phi$ and $U_{4}(\phi)=\frac{1}{8\pi^2}\phi^2(\phi{-}2\pi)^2$ are the sine-Gordon 
and $\phi^4$ parts of the self-interaction potential $U(\phi)$. The sine-Gordon integral can be calculated explicitly with the result
\begin{equation}
    W_{sG} = 16\left[1-\frac{1}{2\cosh^2(2a)}\left(1+\frac{2a}{\sinh(2a)}\right)\right].
\end{equation}
 Unfortunately we were unable to find a closed form of the effective potential coming from the $\phi^4$ part, which we denote as $W_4$. 
 For large positive values of $a\gg 1$, corresponding to a kink - antikink pair,  the effective potential tends to a 
 constant, $W_4\to13.426$. 
 Large negative values of $a\ll-1$ correspond to a `wrongly-ordered' antikink-kink pair separated by a region of the false vacuum. 
 In this case $W_4$ ultimately grows linearly with $|a|$, with gradient $16\pi^2$, since the false vacuum section of length $2|a|$ has 
 the constant energy density of $\epsilon U(-2\pi)=-8\pi^2\epsilon$. Numerical integration for $a\ll -1$ gives an asymptotic 
 form $W_4(a)\approx 157.91 |a|-188.52$, where the slope is equal to $16\pi^2$ as expected.

\begin{figure}[!ht]
\begin{center}
\includegraphics[width=0.75\textwidth]{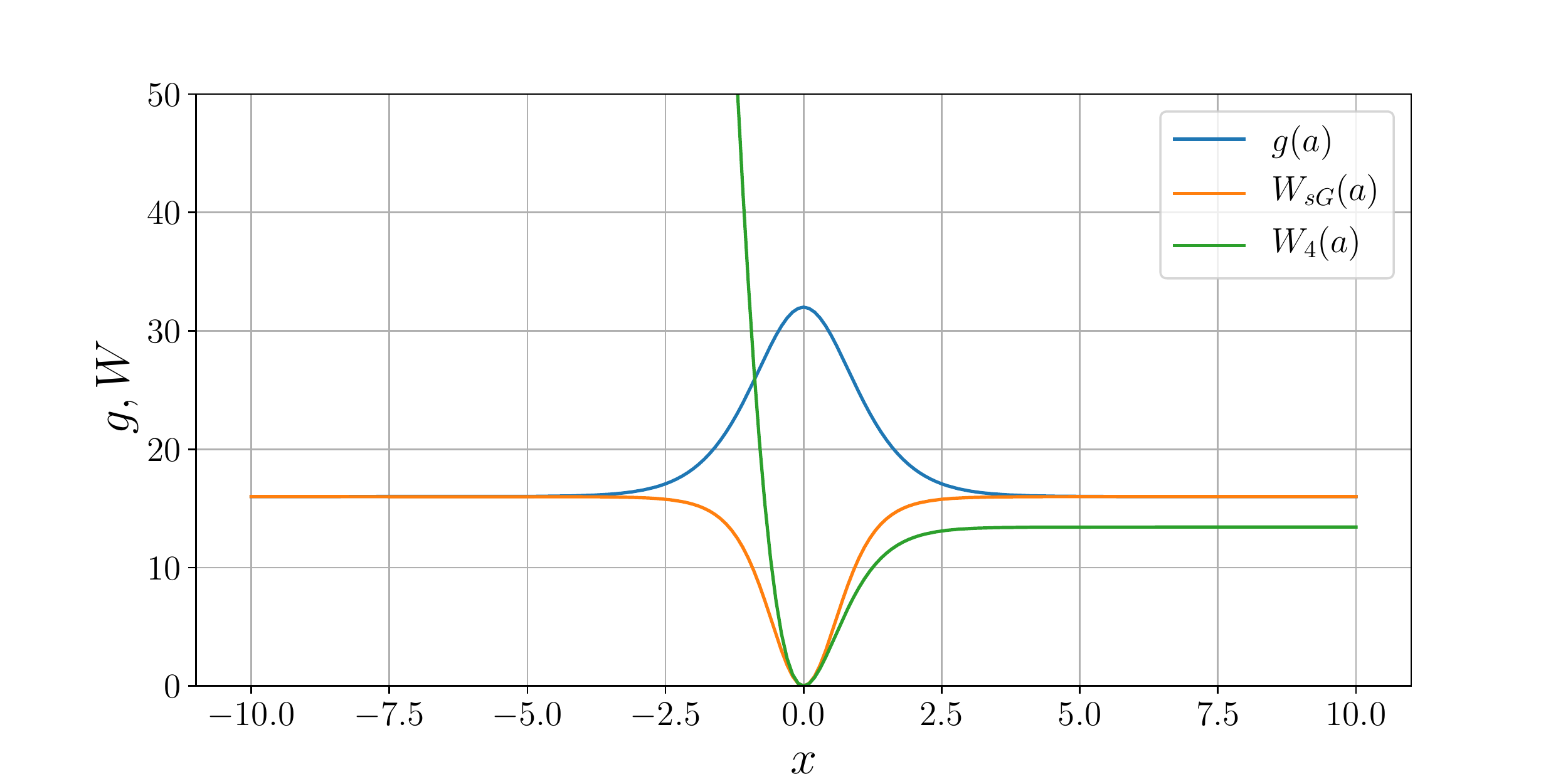}
\end{center}
\vskip -15pt
\caption{Metric and effective potentials for small values of $\epsilon$}
\label{effectiveModel}
\end{figure}

\begin{figure}[!ht]
\begin{center}
\hspace*{-15pt}\includegraphics[width=1.06\textwidth]{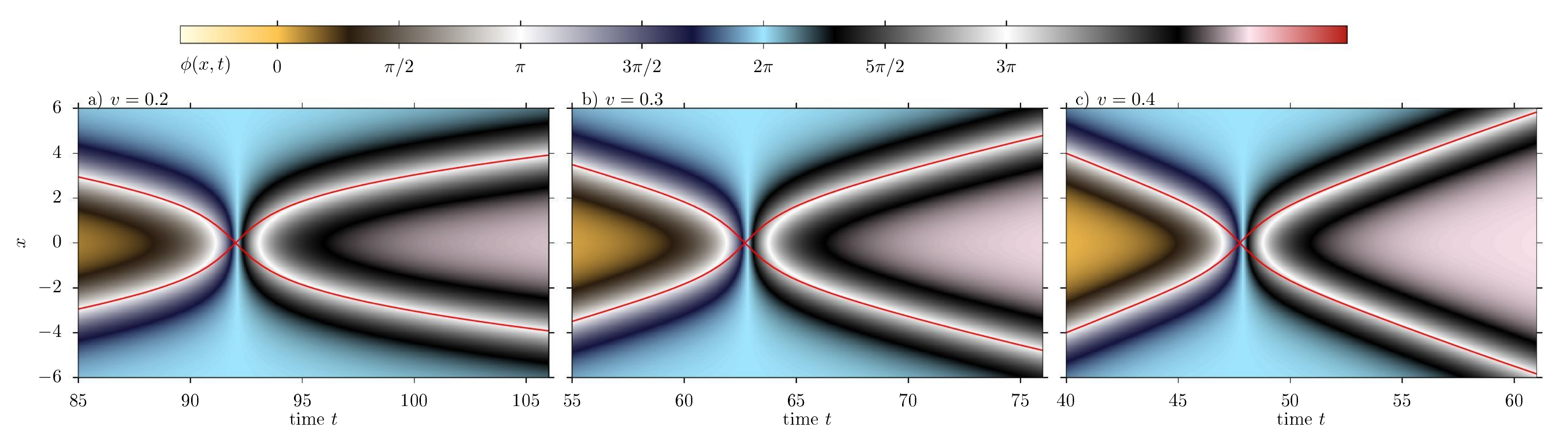}
\end{center}
\vskip -15pt
\caption{Close-up views of the initial collision for the top row of plots of figure \ref{false_collisions}, comparing the full field theory results with the predictions of the effective model, which are shown with continuous red lines.}
\label{mapsEffective_3x1}
\end{figure}

The three functions determining the effective dynamics are shown in figure \ref{effectiveModel}, and trajectories obtained from this model are included as dashed red lines in figure \ref{false_collisions}. The discrepancy is greatest down the right hand column, as expected since the effective model should be most accurate for low velocities. Indeed, the effective model trajectories and those obtained from full simulations are identical within the resolution of the figure for $v=0.2$. Figure \ref{mapsEffective_3x1} zooms in to the region of the first collision for the plots along the top row of figure \ref{false_collisions}, showing that the effective model manages to capture the dynamics even when the kinks collide.

\subsection{Perturbations from $\phi^4$}
As $\epsilon$ increases beyond $0.05$ the false vacuum at $\phi\approx -2\pi$ is lost, 
and after a transitional region 
the pattern of $\bar K K$ collisions comes to resemble the well-known resonant behaviour found in in the $\phi^4$ model
\cite{Anninos:1991un,Campbell:1983xu,Goodman:2005,Makhankov:1978rg,Moshir:1981ja}, which is recovered for $\epsilon=1$.
Provided the initial velocity $v_{i}$ of the colliding kinks is larger than an upper critical velocity
$v_c$, 
they rebound and, while losing some of their velocity, escape back to infinity. Correspondingly the field at the origin reverts to its initial value of $0$, coloured yellow on figure \ref{fullscan}. For $v_i<v_c$, the rebounding kinks from the initial collision no longer have escape velocity and recollide at the centre, after which they generally form
an oscillating bound state, a so-called bion (sometimes also called an oscillon in the literature\footnote{Some authors make a distinction between the two terms, using the word oscillon for objects with well defined frequencies, but we will use the word bion for any form of slowly-decaying oscillating bound state throughout this paper.}),
which slowly radiates the energy away, annihilating into the vacuum $\phi=2\pi$ as $t\to\infty$. This vacuum is coloured blue on figure~\ref{fullscan}. 

However it turns out that
there are `windows' of velocities
 well below $v_c$, within which the recolliding kinks recover their ability to escape. For initial velocities in these ranges, the kinks collide, separate to a finite distance and
turn around to collide a second time, and then escape to infinity. In fact there is a tower of these
`two-bounce' windows of decreasing width, labelled by an integer `window number' $n$
related to the number of the oscillations of the kink's internal mode between the two collisions
\cite{Anninos:1991un,Campbell:1983xu,Goodman:2005}.
This sequence of windows accumulates at the upper critical velocity $v_{c}$ as $n\to\infty$.

These resonance windows are associated with the reversible exchange of energy
between the internal and translational modes of the kinks, the condition for resonance leading to a prediction for the
relationship between the window number $n$,
the time interval between the two collisions $T$, and
the frequency $\omega_1$ of the internal modes of the colliding kinks \cite{Campbell:1983xu}
\be
\omega_1 T = 2\pi n + \delta
\label{res}
\ee
where $n$ counts the number of oscillations of the internal mode between the two collisions and $\delta$ is a phase shift which we fix, as in 
\cite{Campbell:1983xu},
to lie in the range $0\le\delta<2\pi$. For some values of $n$ the energy imparted to the kinks after the second collision, while still high, might not be enough to allow their escape. These correspond to so-called false windows.

\begin{figure}[ht!]
\begin{center}
\setlength{\unitlength}{0.1cm}
\includegraphics[height=.32\textheight, angle =0]{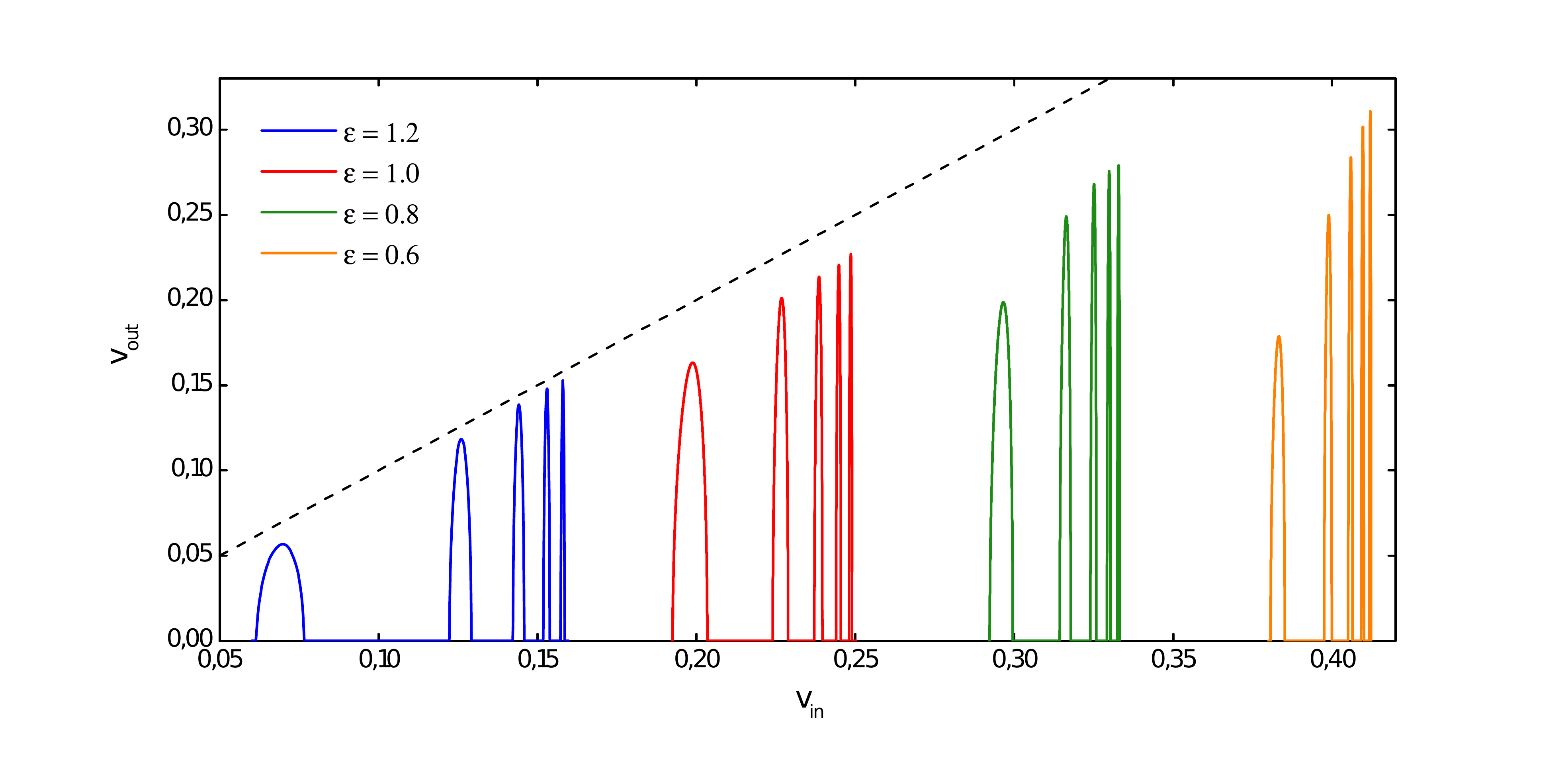}
\end{center}
\vskip -30pt
\caption{Two-bounce windows in regime 1: the final velocity of the kink-antikink pair as a function of the
initial velocity for various values of the deformation parameter.
}
\label{fig5}
\end{figure}

The picture of $\phi^4$-like resonant scattering just described is preserved in our model
provided the deformation parameter $\epsilon$ is not too far
from one. Figure \ref{fig5} shows some typical patterns of two-bounce windows, plotting
the final velocity of the kinks $v_{out}$
as a function of the initial velocity $v_{in}$ inside these windows for various values of $\epsilon$, and
showing that the deformed model \re{lag} still supports sequences of resonance windows
in this regime. Further three and higher
bounce windows exist at the edges of each two-bounce window,
but are not shown in the plot; we will not explore this aspect of the models in this paper.

\begin{figure}[!ht]
\begin{center}
\setlength{\unitlength}{0.1cm}
\includegraphics[width=.6\textwidth, angle =0, trim = 60 15 60 60, clip=true]{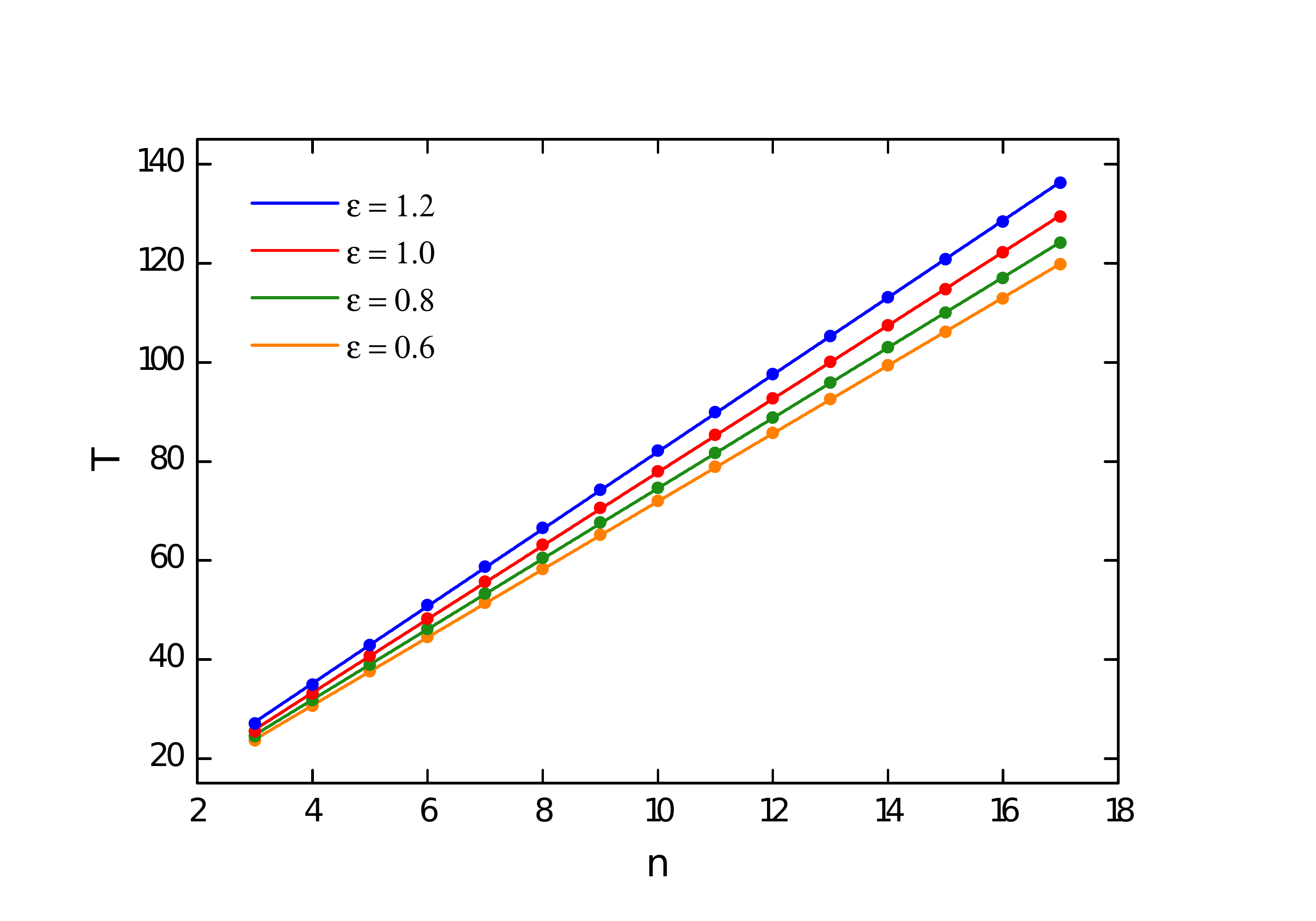}
\end{center}
\vskip -15pt
\caption{The time $T$ between the two collisions as a function of the window number $n = 3\dots 17$ for two-bounce
resonance windows with $\epsilon = 1.2,~1,~0.8$ and $0.6$.
}
\label{TN}
\end{figure}

Figure \ref{TN} plots the inter-collision times for these two-bounce windows against the window number $n$, showing that the
satisfy the expected linear relationship (\ref{res}), as for the $\phi^4$ model. Fitted slopes and intercepts of the lines on the figure are given in table~\ref{slopes}. The second and fifth columns show reasonable agreement, as predicted by~(\ref{res}). 

\begin{figure}[ht!]
\begin{center}
\setlength{\unitlength}{0.1cm}
\hspace{-0.45 cm}
\includegraphics[width=\textwidth, angle =0]{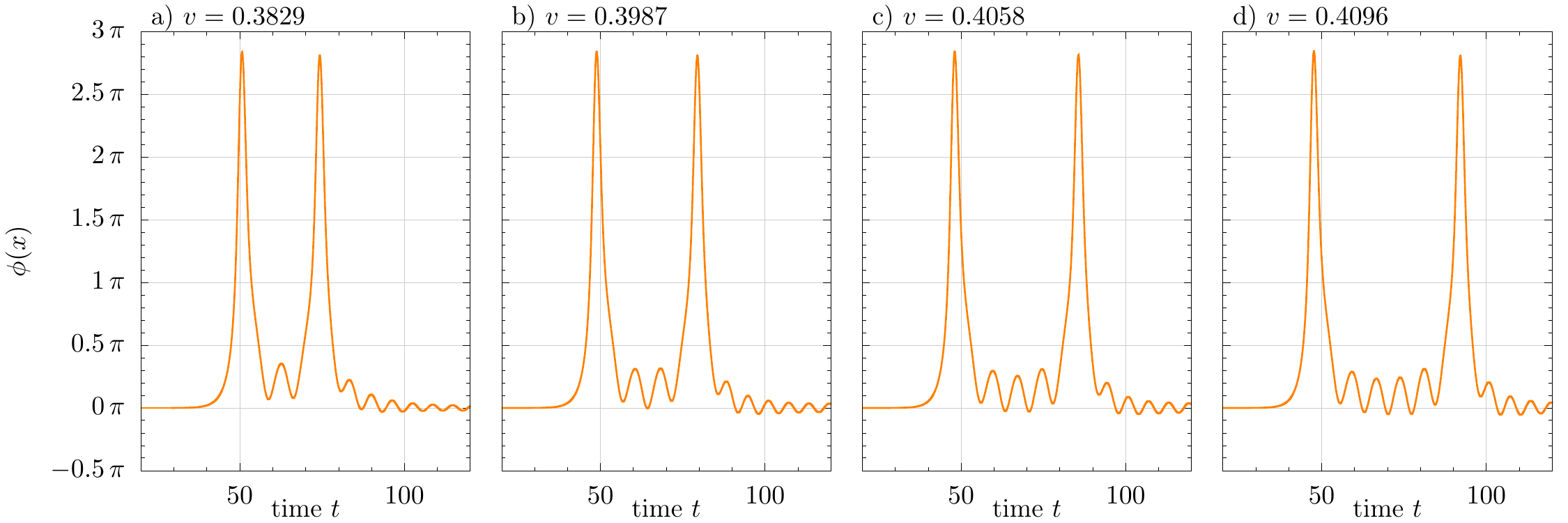}
\end{center}
\vskip -15pt
\caption{Field value at the collision centre as a function of time, showing the first four two-bounce resonance windows 
for $\epsilon=0.6$. For this value of $\epsilon$, $\omega_1=0.939337$ and $2\pi/\omega_1=6.68896$. These plots correspond to window numbers
$n = 3$, $4$, $5$ and $6$ in the formula (\ref{res}).
}
\label{centre_view_eps_06}
\end{figure}

\begin{table}[ht]
\centering
{\small
\begin{tabular}{lcccccc}
\hline
&$\epsilon$&\mbox{slope~}&\mbox{intercept}&$\omega_1(\epsilon)$&$2\pi/\omega_1(\epsilon)$\\
\hline\hline
&\texttt{1.2}&\texttt{7.7907 } &\texttt{3.9929}&\texttt{0.82163761 }
&\texttt{7.647149146}\\
&\texttt{1.0}&\texttt{7.4153 } &\texttt{3.6165}&\texttt{0.86602540 }
&\texttt{7.255197456}\\
&\texttt{0.8}&\texttt{7.1071 } &\texttt{3.4486}&\texttt{0.90521781 }
&\texttt{6.941075670}\\
&\texttt{0.6}&\texttt{6.8621 } &\texttt{3.2652}&\texttt{0.93933673 }
&\texttt{6.688959440}\\
\hline 
\end{tabular}
}
 \caption{Linear fits for the lines on figure \ref{TN}, with the slopes compared with the predictions of the resonance condition (\ref{res}). The values of $\omega_1(\epsilon)$ follow from the (numerical except for $\epsilon=1$) solution of the eigenvalue problem (\ref{pert}), as described in section \ref{SpectStruct}.}
 \label{slopes}
\end{table}
We conclude that the pattern of these windows can be explained by the same give-and-take energy transfer 
mechanism as in the original $\phi^4$ theory. 
The working of the resonance mechanism is clearly seen in figure \ref{centre_view_eps_06}, which shows field evolution at the origin inside the
first four two-bounce windows for $\epsilon=0.6$.

As the
deformation parameter $\epsilon$ decreases from $1$, the upper critical velocity $v_c$ initially increases
almost linearly, as can be seen in figures \ref{fullscan} and \ref{fig5}. At the same time  
the associated windows, and the gaps between them, become narrower. 
    
\begin{figure}[ht]
\begin{center}
\setlength{\unitlength}{0.1cm}
 \includegraphics[width=1\textwidth, angle =0]{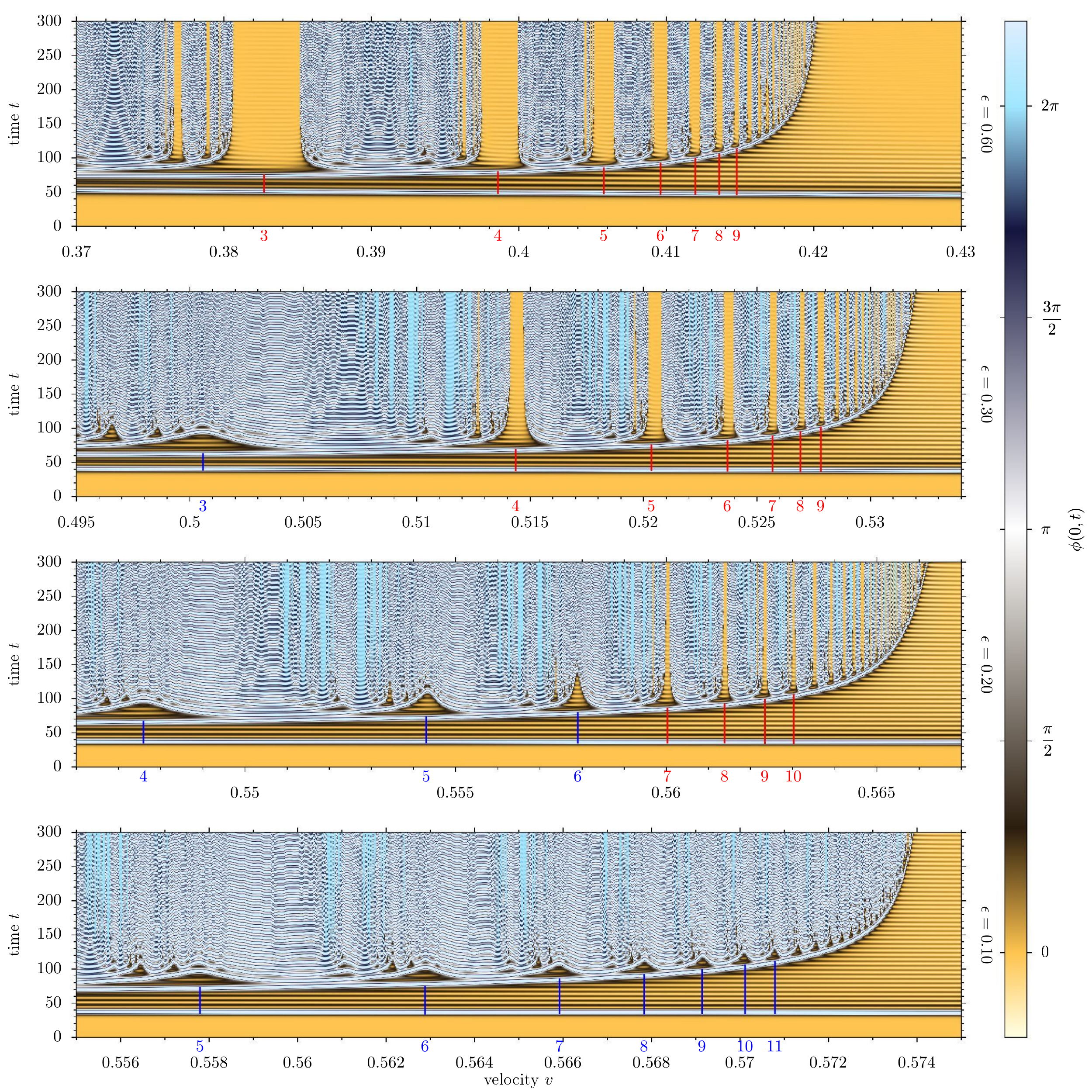}
\end{center}
\vskip -15pt
\caption{
Kink-antikink resonance collisions in the model \re{lag} for four values of the
parameter $\epsilon$ in regime 1. The plots show the field value at the
collision centre as a function of initial velocity $v$ and time $t$ from the start of the simulation. As $\epsilon$ decreases, the windows move to higher velocities, and some change from true windows to false windows. The window numbers are shown in red and blue below the plots; blue for false windows, and red for true ones.
}
\label{ClosedWindows}
\end{figure}

The maximal value of the upper critical velocity $v_{c} = 0.5803$ occurs for
$\epsilon = 0.13$; the corresponding eigenvalue of the internal  mode is $\omega\approx 0.975$.
Evidently, as the frequency of the internal mode is approaching the continuum threshold,
an excitation of the internal mode may also affect the modes of continuum leading to energy loss due to radiation.
This effect can destroy the fine mechanism of the reversible energy exchange in kink-antikink collisions.
Indeed, we observe that the structure of resonance
windows in $K\bar K$ collisions is damaged as $\epsilon$ decreases.

\begin{figure}[ht]
\begin{center}
\setlength{\unitlength}{0.1cm}
\includegraphics[width=1\textwidth, angle =0]{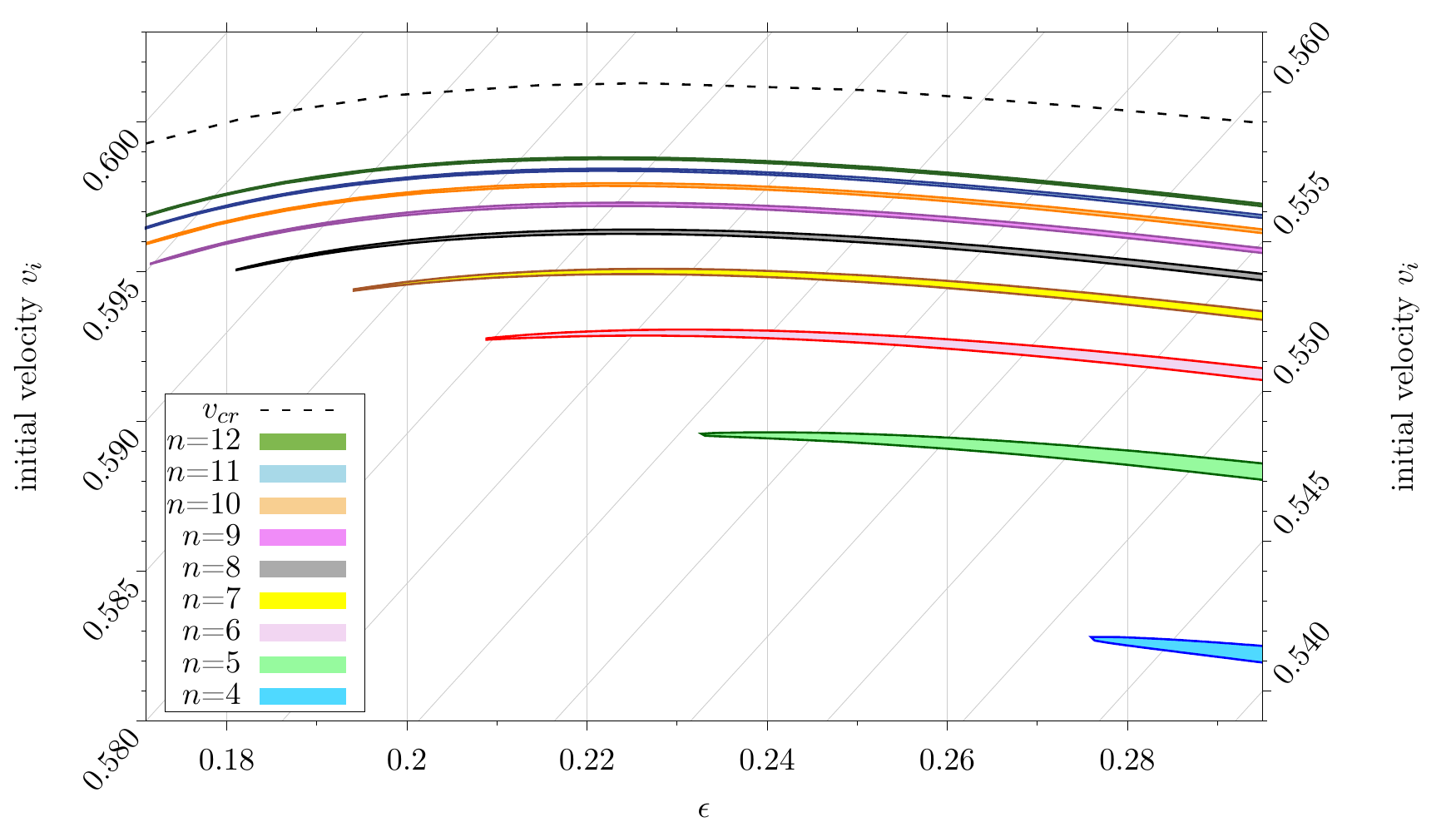}
\end{center}
\vskip -15pt
\caption{
The closing two-bounce windows as a function of both $\epsilon$ and initial velocity $v_i$ in a sheared coordinate system to reduce the high inclination of the structure.} 
\label{ClosedWindows2}
\end{figure}

Surprisingly, the first two-bounce window that becomes false, at about $\epsilon=0.35$, is the one which was initially the widest, corresponding to the lowest number $n$ of oscillations of the internal mode. Figure \ref{ClosedWindows} shows the
loss of true windows in action. In the top plot, with $\epsilon=0.6$, all the windows seen at $\epsilon=1$ are still open. 
As $\epsilon$ decreases through $0.35$, the smallest-velocity window, for which $n=3$, becomes false, as seen in the second plot, taken at $\epsilon=0.3$. By the time $\epsilon$ has decreased to $0.2$, shown in the third plot, three more windows have become false. Finally the bottom plot shows that all resonance windows have closed at $\epsilon=0.1$.
To show the pattern in more detail we located the two-bounce windows for $0.17<\epsilon<0.295$ and plotted them in figure \ref{ClosedWindows2}. Because the upper critical velocity as well as the whole structure is a steep function of $\epsilon$ in this range ($\Delta v/\Delta \epsilon\approx 0.33064$) we used an affine transformation to reduce the shear effect and kept $\epsilon$ unchanged. The figure shows clearly that the windows with smaller values of $n$ close first as $\epsilon$ decreases.

\section{The second transition: the emergence of the double kink}\label{sec:Regime2}

\subsection{Multiple bound modes}
\begin{figure}[!ht]
\begin{center}
\includegraphics[width=1.0\textwidth]{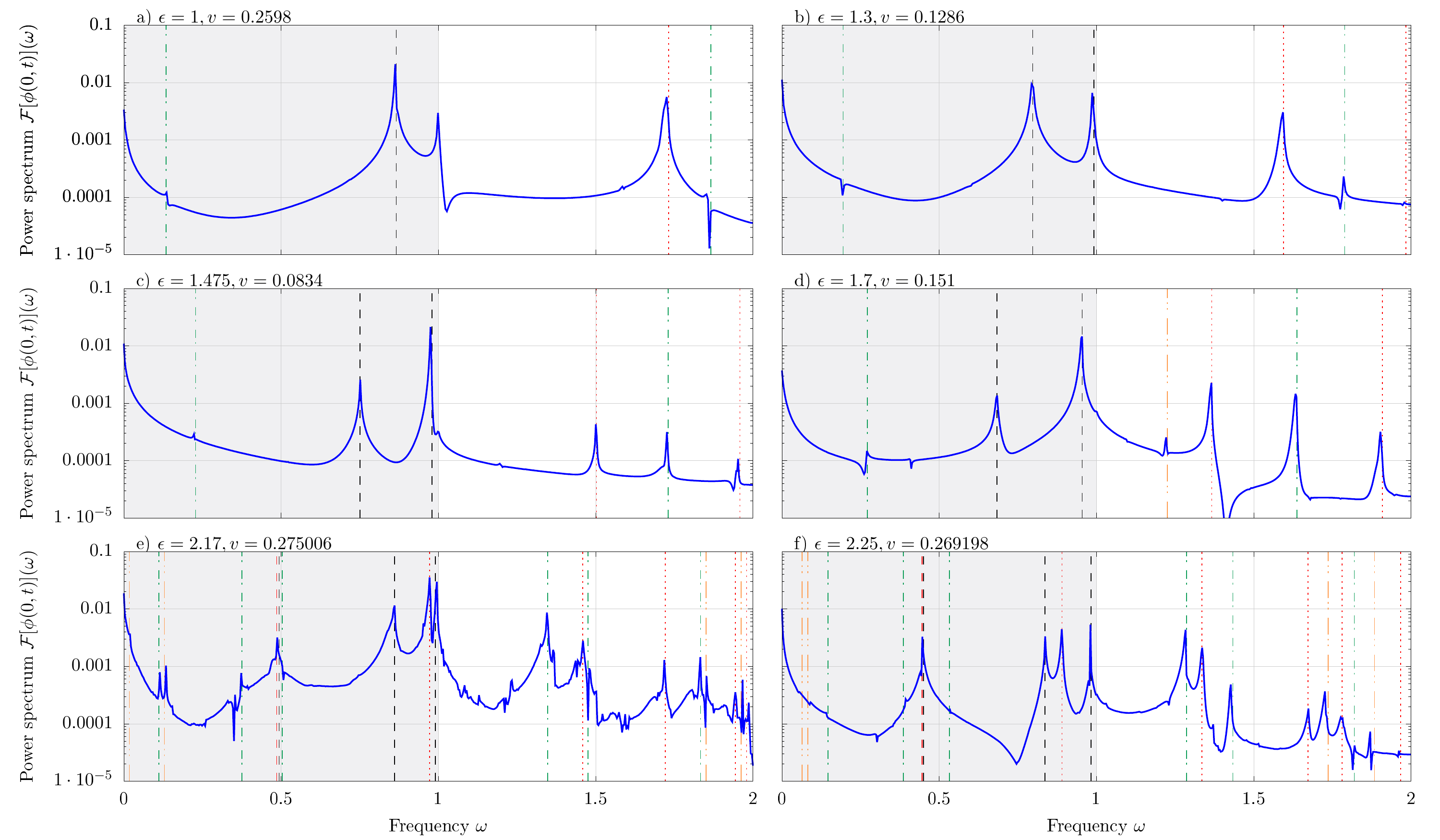}
\end{center}
\caption{Power spectra of fluctuations at the origin after a kink-antikink collision
for various values of $\epsilon$ from $1$ to $2.25$, showing 
signals of the extra internal kink modes which emerge as $\epsilon$ increases past $1$ and then $2$. The speeds of the incident kink and antikink, shown above each subplot, is chosen in each case to be just larger than $v_c$ for that value of $\epsilon$. Dashed black lines
show the frequencies of the internal modes $\omega_1$, $\omega_2$ and so on; red dotted lines are integer multiples of these frequencies; green dot-dashed lines are at $|\omega_i\pm\omega_j|$, $i\neq j=1,2,3$; and orange dot-dashed lines at $|2\omega_1\pm\omega_i|$, $i=2,3$.}
\label{fig16}
\end{figure}

\begin{figure}[!ht] 
\begin{center}
\includegraphics[width=1\textwidth]{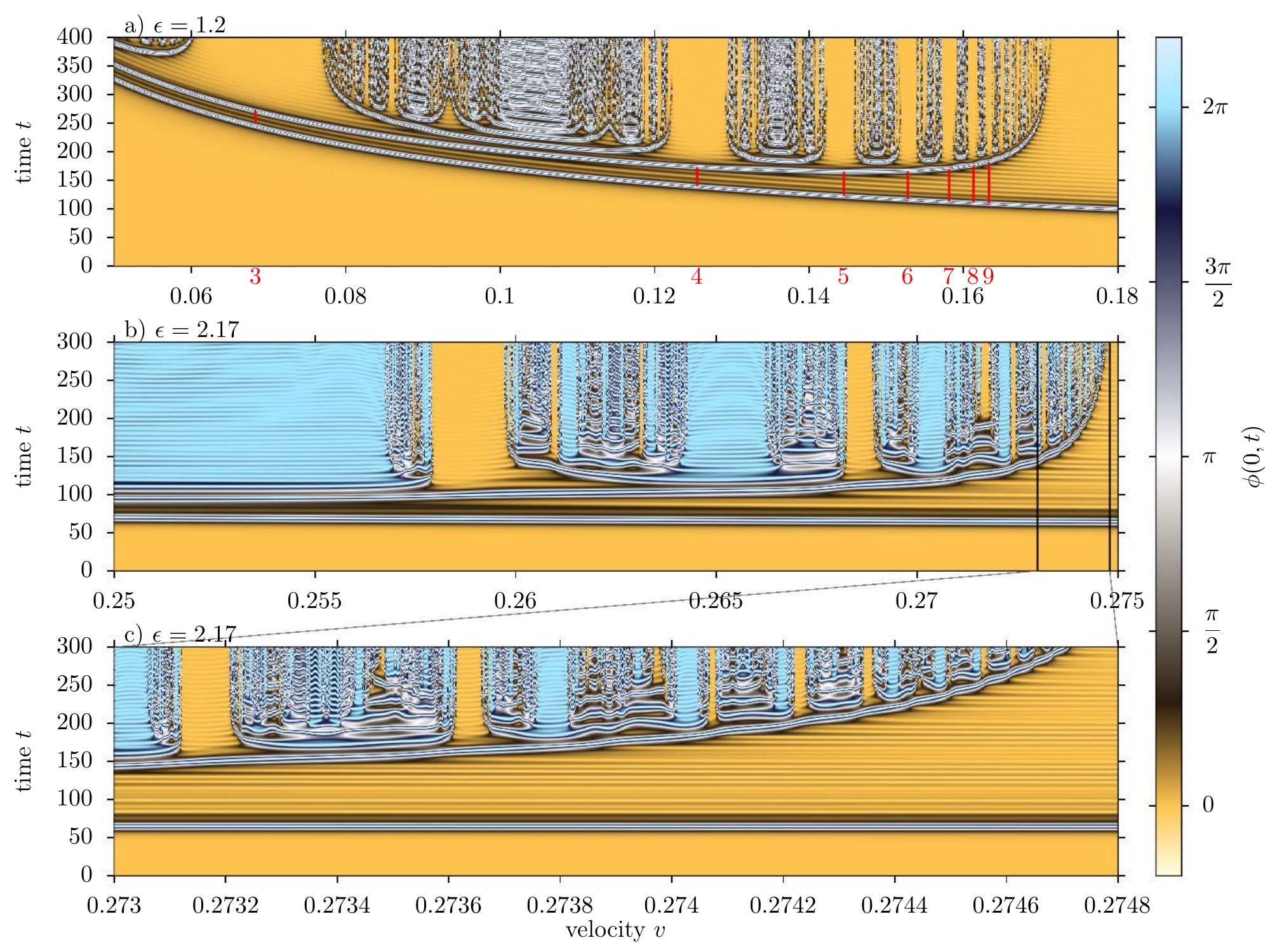}
\end{center}
\vskip -15pt
\caption{Field at the centre for $\epsilon=1.2$ (a) and $\epsilon=2.17$ (b), (c). Subplot (a) shows a regular structure of $\phi^4$-like windows labelled by the oscillation number $n$ of the first internal mode, as seen for $\epsilon\le 1$ in
figure \ref{ClosedWindows}. Subplots (b) and (c) are taken for a value of $\epsilon$ for which the window structure is much more chaotic. The pattern includes a number of the blue `pseudowindows' discussed more in the main text. 
}
\label{velocityscanx3}
\end{figure}

While the deformation parameter $\epsilon$ remains below 1, there is just one internal mode of the kink. 
As  $\epsilon$ increases past this value, a second internal mode, with frequency $\omega_2$, 
emerges from the continuum, as shown in figure \ref{spectral_structure}. The presence  of this can be seen numerically in the power spectra of the field at the origin after a kink-antikink collision, shown in figure~\ref{fig16}. The extra internal mode complicates the pattern of reversible energy exchange between the translational
mode of the kinks and the internal modes, but while $\epsilon$ remains below about $1.4$,  
the frequency of the second mode $\omega_2$ is not very far below the mass threshold, 
and the leading role in the resonance scattering mechanism still
belongs to the lowest frequency internal mode $\omega_1$.
The structure of $\phi^4$-like
escape windows labelled by the oscillation number $n$ of this mode survives, as seen in
figure \ref{velocityscanx3} (a) and the blue curves in figures \ref{fig5} and \ref{TN}. As
$\epsilon$ increases through $1.25$ these windows start to close, again starting 
from the smallest values of $n$, and by $\epsilon\approx 1.5$ they have all gone. 
The resonance condition is no longer given by the simple relation \re{res}, and
the interplay between the two internal modes becomes more important as $\epsilon$ increases further and the second mode becomes well separated from the continuum. 
The upper critical velocity $v_{c}$ also begins to increase as $\epsilon$ becomes larger than 1.5, as seen in figure \ref{fullscan}.

\begin{figure}[!ht]
\begin{center}
\includegraphics[width=0.82\textwidth]{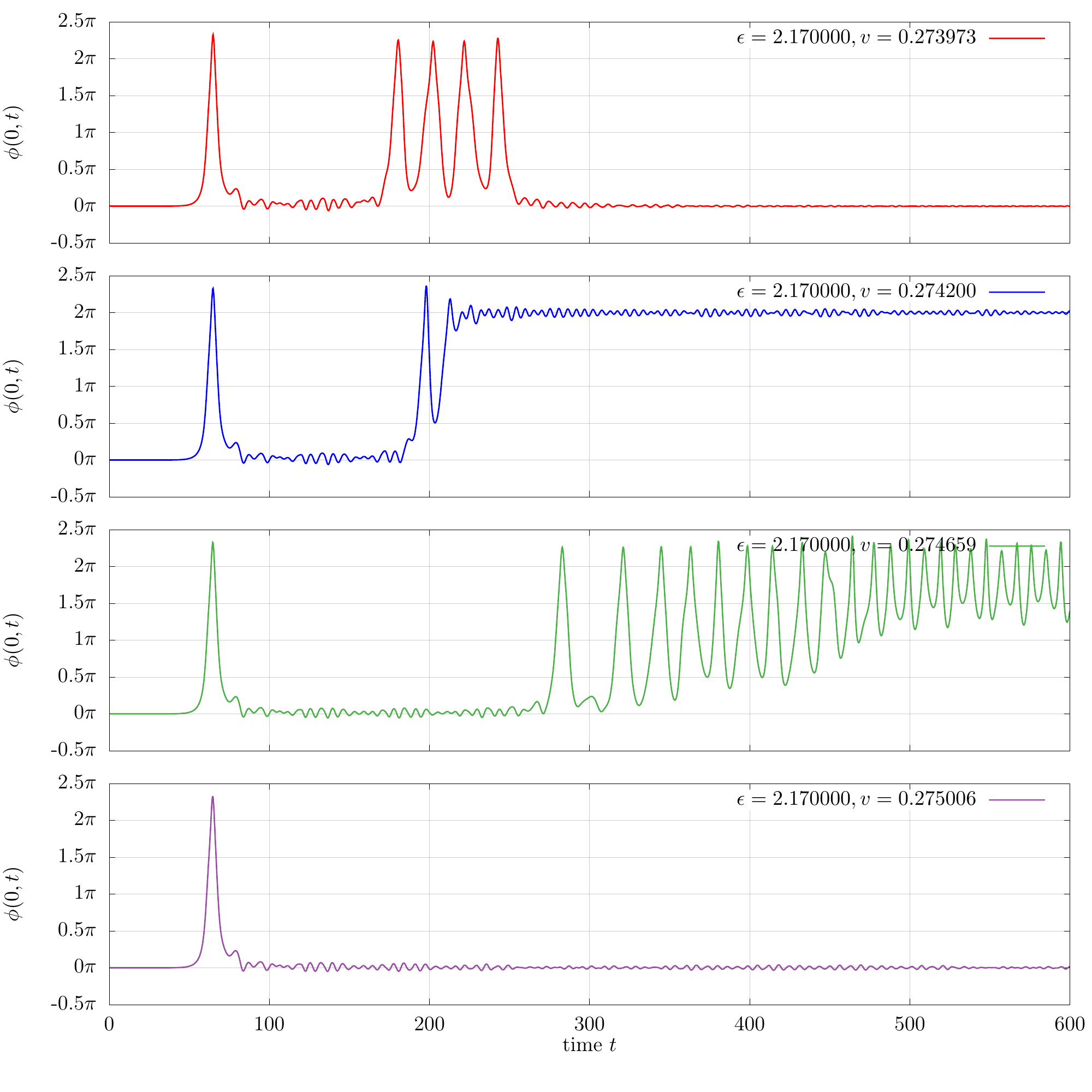}
\end{center}
\vskip -20pt
\caption{Evolution of the field at $x=0$ for four values of the initial velocity with $\epsilon=2.17$, $\beta=1.19$
showing a true window, a pseudowindow, an annihilation to a centrally-located bion, 
and kink-antikink escape.
The presence of higher modes can be seen in the more irregular oscillations of the field after the first collision compared to figure \ref{centre_view_eps_06}. Longer-time views of the full spacetime evolution of the field for these collisions are shown in figure \ref{mapsCriticalVelocity} below.}
\label{fig15}
\end{figure}

\bigskip

\begin{figure}[!ht]
\begin{center}
\includegraphics[width=0.97\textwidth]{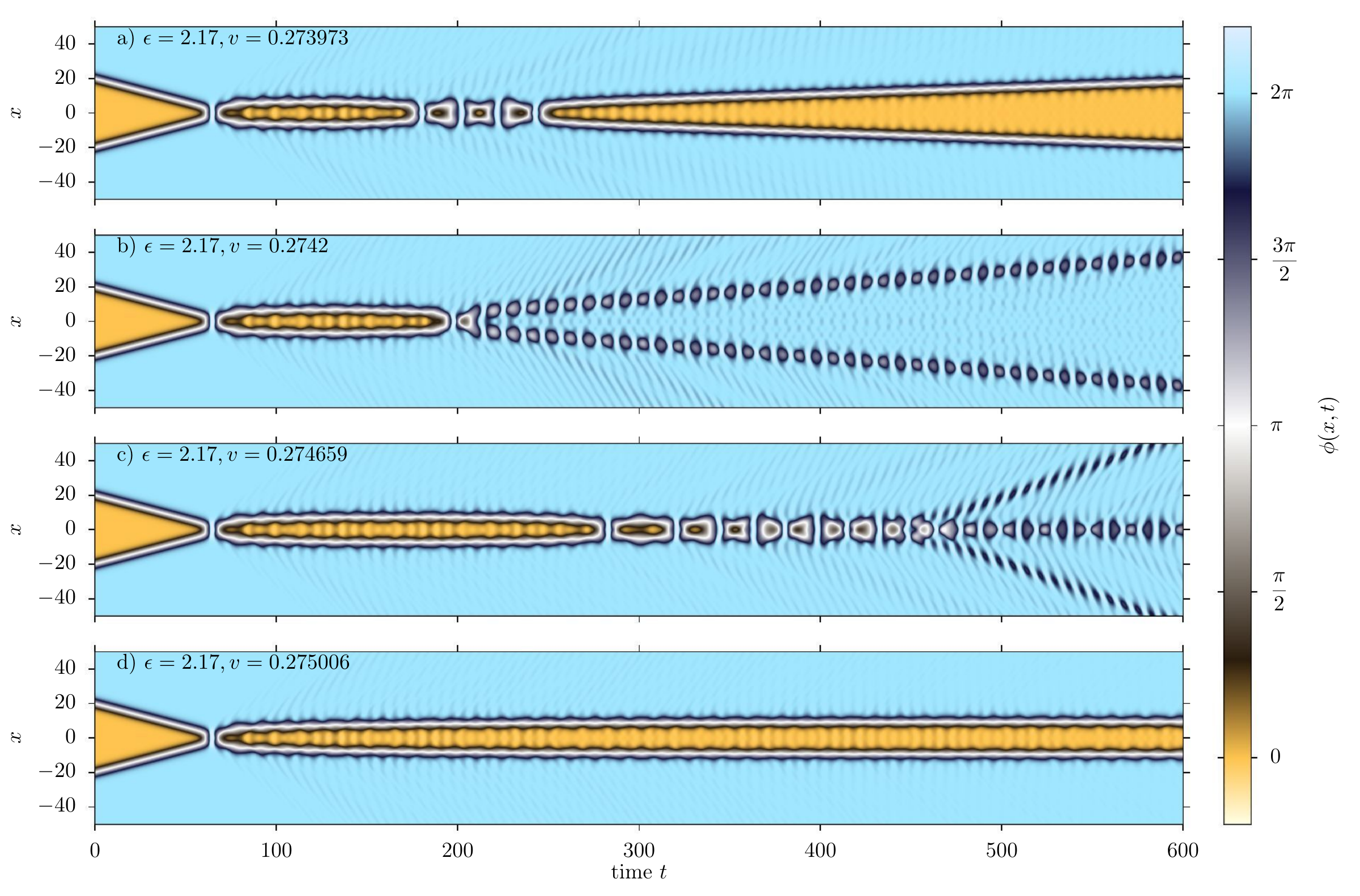}%
\end{center}
\vskip -20pt
\caption{Spacetime plots of the four collisions of figure \ref{fig15}. The second plot shows the emmission of a pair of bions at $t\approx 200$, which rapidly transport energy away from the central region.
The centrally-located bion in the third plot emits a burst of radiation at $t\approx 460$; this is an example of the stacatto radiation phenomenon discovered in \cite{Dorey:2019uap}.}
\label{mapsCriticalVelocity}
\end{figure}

Increasing $\epsilon$ further yields more complicated chaotic dynamics, and the sporadic reappearance of scattering windows below the upper critical velocity. The third internal
mode appears in the spectrum of linearized perturbations while the frequencies of the first and the second modes
continue to decrease. Figures~\ref{velocityscanx3} (b), (c)
show the window structure at $\epsilon=2.17$, while figures \ref{fig15} and \ref{mapsCriticalVelocity} show some typical collision processes at the same value of $\epsilon$. 
The windows again accumulate as the critical velocity $v_c$ is approached, but the pattern is much less regular than the $\phi^4$-like resonant $\bar K K$ collisions considered thus far. 
The escape windows (indicated by the yellow colour) become less symmetric, as if one side of the window has merged with a false window. 
In addition, a set of light blue regions that we will call `pseudowindows' appears.
Their blue colour indicates that the centre field after collision is flipped into the $\phi=2\pi$ vacuum. Topologically, this corresponds to an annihilation: the value of the
field at the origin in the final state is equal to its value at infinity. 
However, instead of there being slowly-decaying bion at the centre (signalled by white-grey-black stripes on figure \ref{velocityscanx3}), as formed  
in a $\phi^4$-like $\bar K K$ capture, two bions are ejected in opposite 
directions, rapdily transporting most of the energy away from the origin and
leaving the centre field oscillating around the $2\pi$ vacuum with a relatively small amplitude.
The collisions shown in the
second and third plots of figures \ref{fig15} and \ref{mapsCriticalVelocity} show this distinction
clearly: both are examples of antikink-kink capture, but in the second plot a pair of bions is emitted and the central
field relaxes quickly to $2\pi$, while in the third plot a bion remains at the origin and the relaxation is much slower. Note that 
the concept of a pseudowindow relies on field measurements being taken at intermediate time scales, longer than the collision time but shorter than the time required for the radiative decay of a bion. At larger time scales the field at any fixed location, including the origin, will relax to one of the vacuum values, and so regions exhibiting white-grey-black stripes in our plots will ultimately revert to the same blue colour as the pseudowindows. Nevertheless, the extremely long lifetime of the bion means that the relevant time-scales are well separated, making the pseudowindows clearly visible.
As discussed in the next section, 
more examples of this phenomenon can be observed for higher values 
of $\epsilon$, closer to $\epsilon_{cr}$. In fact, for very small values of $\epsilon$  some narrow pseudowindows also appear, as can be
seen in figure~\ref{ClosedWindows}.

\begin{figure}[!ht]
\begin{center}
\includegraphics[width=1.0\textwidth]{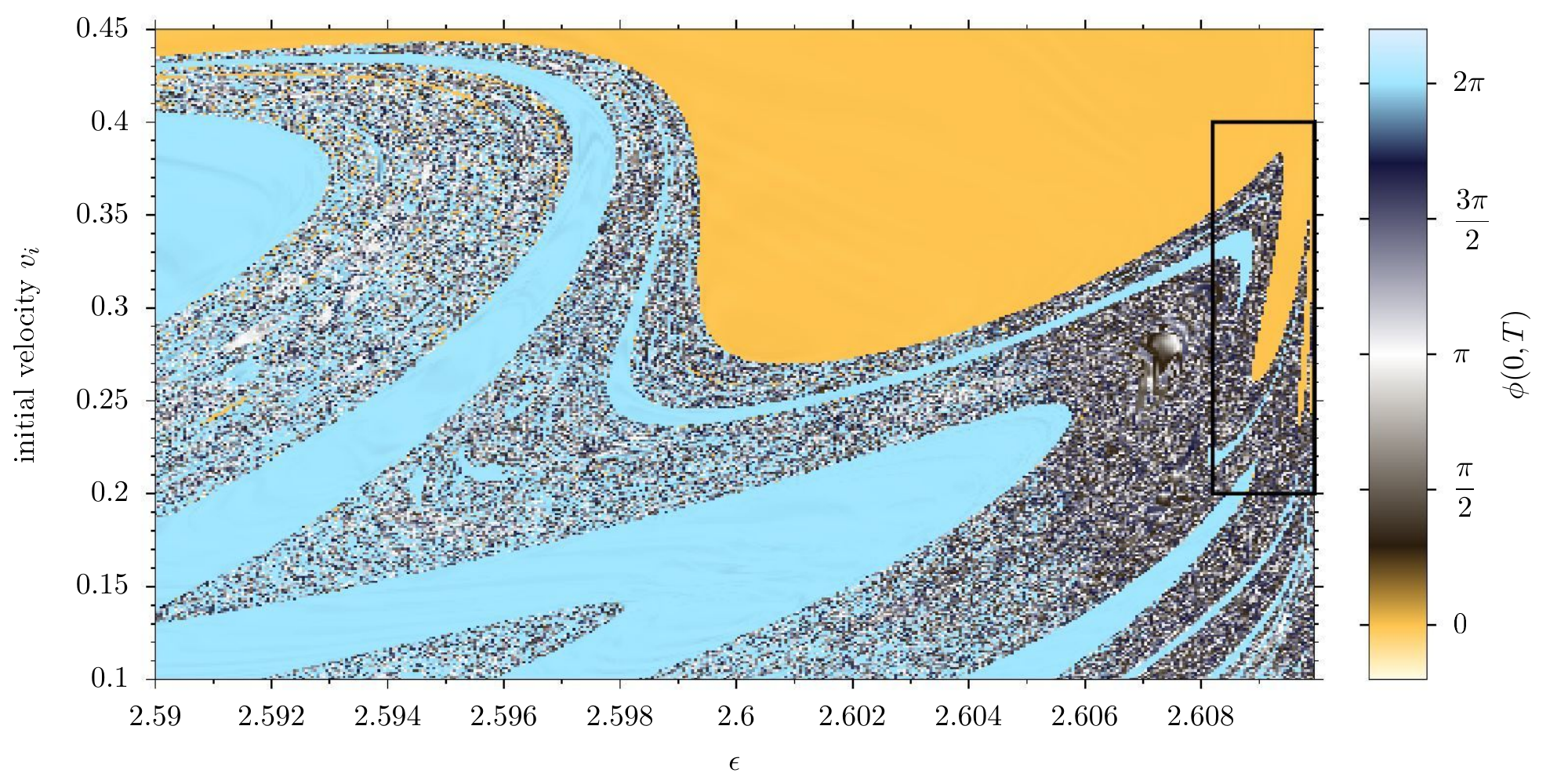}
\end{center}
\caption{The post-collision value of the field at the collision centre as a
function of initial velocity of the colliding kinks, and the deformation parameter $\epsilon$ near the critical value. The field value was measured $300$ units of time after the intersection of the initial trajectories, that is $300+x_0/v_i$ units of time after the start of the simulation. The black rectangle to the right indicates the near-critical region plotted in figure \ref{nearcritscan_zoom}.}
\label{nearcritscan}
\end{figure}

The full scan shown
in figure \ref{fullscan} and the closer zoom for $\epsilon$ near to $\epsilon_{cr}$ plotted in figure \ref{nearcritscan} 
show that the critical velocity oscillates as $\epsilon$ increases beyond $1.5$. A possible explanation for this effect is that the local maxima of the escape velocity correspond to maximal loss of the energy due to radiation. This can happen when two conditions are fulfilled: (i) a large amount of energy is transferred to one of the internal modes; (ii) this mode quickly radiates the energy away via the coupling to the continuum. Indeed,  the modes closest to the mass threshold radiate the most. 
One of the modes is also
exactly at the threshold for $\epsilon=0$. However, this value corresponds to the integrable sine-Gordon model. Hence, the threshold  mode is decoupled from the solitonic degrees of freedom, and cannot be excited during collisions. As a matter of fact, the first local maximum of the critical velocity is for $\epsilon=0.131$ for which the internal mode reaches $\omega=0.9951$. Similar values of frequencies of odd modes are reached for two more local maxima (Table \ref{extrema}).

\begin{table}
{\small
\centering
\begin{tabular}{lcccccc}
\hline
&$\epsilon$&$\omega_1$&$\omega_2$&$\omega_3$&$\omega_4$&$\omega_5$\\
\hline\hline
Maxima:\\
&\ \ \ \ \texttt{0.131} \ \ \ \ & \textcolor{blue}{\texttt{0.9951}}& \\
&\texttt{2.171}& \texttt{0.4931}& \texttt{0.8606}& \textcolor{blue}{\texttt{0.9927}}& \\
&\texttt{2.597}&  \ \ \ \texttt{0.0906} \ \ \ &  \ \ \ \texttt{0.6284} \ \ \ 
&  \ \ \ \texttt{0.7679} \ \ \ &  \ \ \ \texttt{0.9057} \ \ \ & \ \ \ \textcolor{blue}{\texttt{0.9978}}\ \ \  \\

Minima:\\
&\texttt{1.471}& \textcolor{red}{\texttt{0.7522}}& \texttt{0.9796}& \\
&\texttt{2.444}& \texttt{0.3132}& \textcolor{red}{\texttt{0.7565}}& \texttt{0.9305}& \\
&\texttt{2.601}& \texttt{0.0755}& \texttt{0.6200}& \textcolor{red}{\texttt{0.7513}}& \texttt{0.8868}& \texttt{0.9896} \\
\hline 
\end{tabular}
}
 \caption{Spectral structure at extreme points}\label{extrema}
\end{table}

Another intriguing observation is that the local minima of the critical velocity correspond to the values of $\epsilon$ for which one of the frequencies reaches the value $3/4$ (within 1\% error). This might indicate that some $3:4$ resonance with the threshold frequency 
plays an important role in the collision process. 
Note that this resonance can happen for both even and odd modes, whereas the maxima are only for the odd modes. This may mean that some higher nonlinear term  (presumably fourth order) is responsible for the 
minima of the critical velocity.

\subsection{Double kink collisions near the critical value of $\epsilon$}
For $\epsilon>2$, 
a false vacuum appears at $\phi=\pi$
between the true vacua at $\phi=0$  and $\phi=2\pi$. 
The presence of this false vacuum deforms the kink, splitting it into two smaller subkinks, 
which become more and more separated as $\epsilon\to\epsilon_{cr}$,
as illustrated in figure \ref{profiles}. The eigenvalue of the first internal mode $\omega_1$ rapidly approaches zero as
this mode smoothly  transforms into an antisymmetric linear combination of the translational 
modes of the subkinks, as explained in section \ref{doublekinkoscillation}. Similarly, a symmetric combination of these modes corresponds to the translational mode $\omega_0$. 

\begin{figure}[!ht]
\begin{center}
\includegraphics[width=1.0\textwidth]{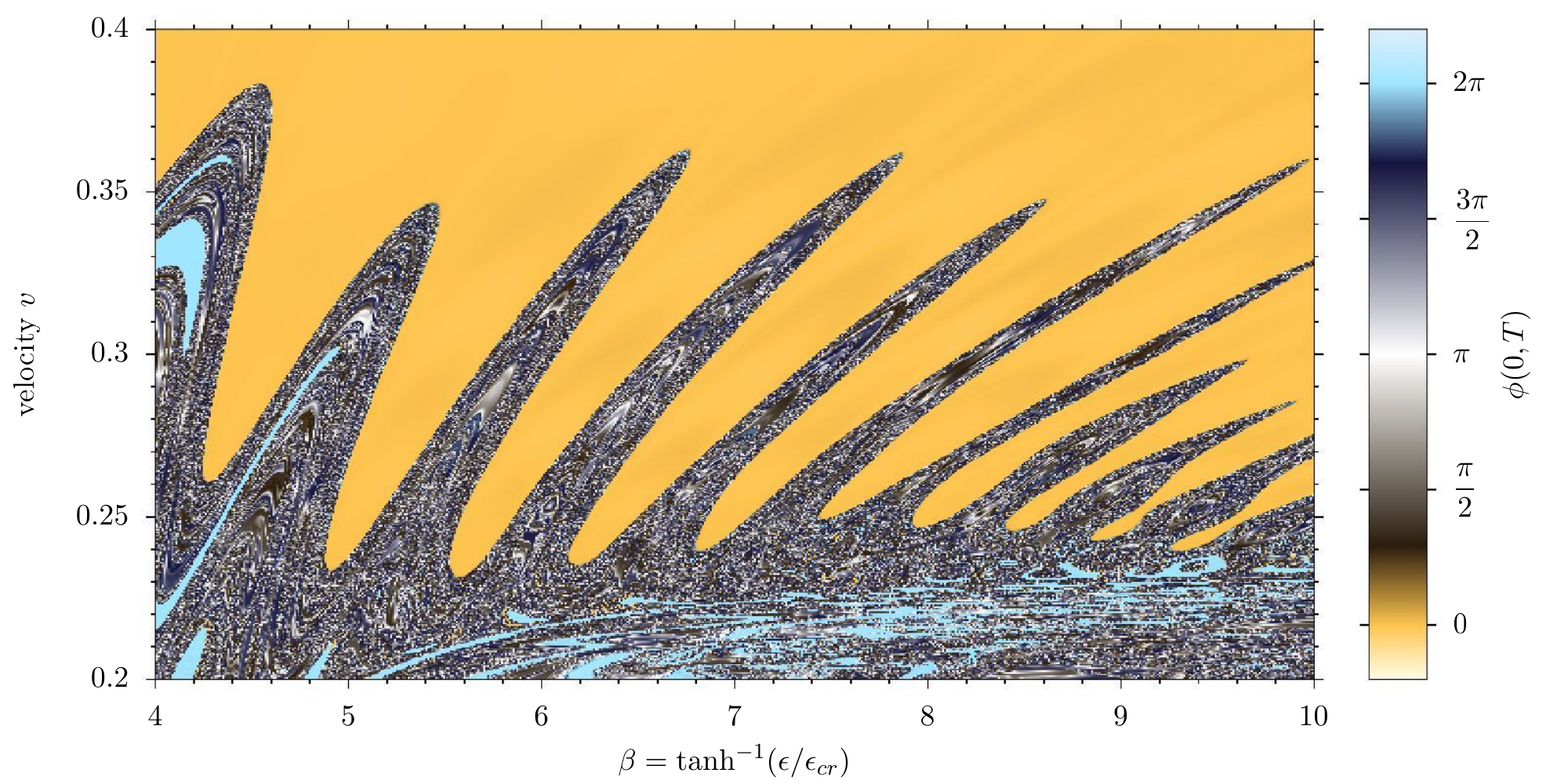}
\end{center}
\vskip -15pt
\caption{The near-critical region of figure \ref{nearcritscan}. The range $4\dots 10$ of the horizontal parameter $\beta=\tanh^{-1}(\epsilon/\epsilon_{cr})$ corresponds to $\epsilon$ running from 
$2.608195$ to $2.609946$. 
For these near-critical values of $\epsilon$ the antikink and kink have increased widths, and so for this plot they were placed further apart at $t=0$, at $x=-40$ and $x=40$. The dangers of placing the initial kinks and antikinks too close together can be seen by examining figure \ref{mapsPressurec} below.
}
\label{nearcritscan_zoom} 
\end{figure}

\begin{figure}[!ht]
\begin{center}
\includegraphics[width=1\textwidth]{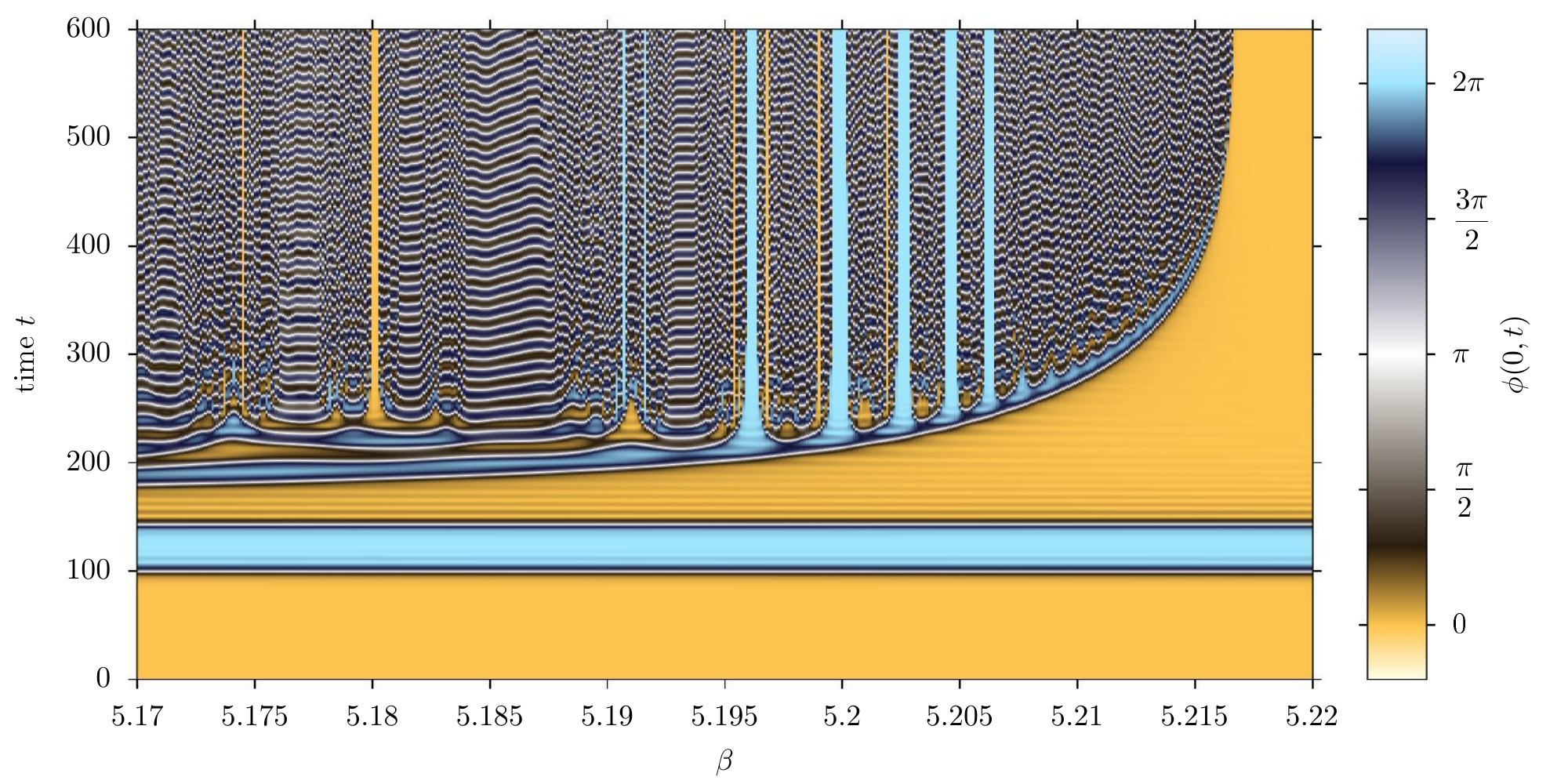}
\end{center}
\vskip -15pt
\caption{Field at $x=0$ for initial velocity $v=0.3$ and a range of $\beta$ cutting into the right-hand 
edge of the second `porcupine spine' in figure \ref{nearcritscan_zoom}, showing both windows 
and pseudowindows.}
\label{ScanBeta1D_c}
\end{figure}

The oscillations of the critical velocity 
become more extreme as $\epsilon_{cr}$ is approached,
their `waves' ultimately overhanging in a series of spines as $\epsilon$ approaches $\epsilon_{cr}$, 
as seen in figure~\ref{nearcritscan_zoom}, where for better resolution we used $\beta=\tanh^{-1}(\epsilon/\epsilon_{cr})$ instead of $\epsilon$ for the horizontal scale.
In preparing this figure we measured the central field, as for our other scans,
at a time $T=300$ after the initial collision. Especially for larger values of $\beta$, the 
final value of the field at the origin may not have settled down by this time, and so we made some
lower-resolution scans with longer evolution times. These showed that the yellow spaces between the 
bases of the spines have a tendency to close with time, but the further away 
from the base, or the smaller the value of 
$\beta$, the closer the spines shown in figure \ref{nearcritscan_zoom} were to their asymptotic forms.
The white-grey-black regions of these spines indicate points where
the incident $\bar KK$ pair annihilate to a centrally-located bion. Within the spines there are also 
(yellow) windows and (light blue) pseudowindows forming meandering stripes, where instead of the 
centrally located bion either a $\bar KK$ pair or a pair of escaping bions is 
produced. A cross-section of the right-hand edge of
one of these spines, exhibiting both windows and pseudowindows, is shown in figure \ref{ScanBeta1D_c}.

\begin{figure}[!ht]
\begin{center}
\includegraphics[width=1\textwidth]{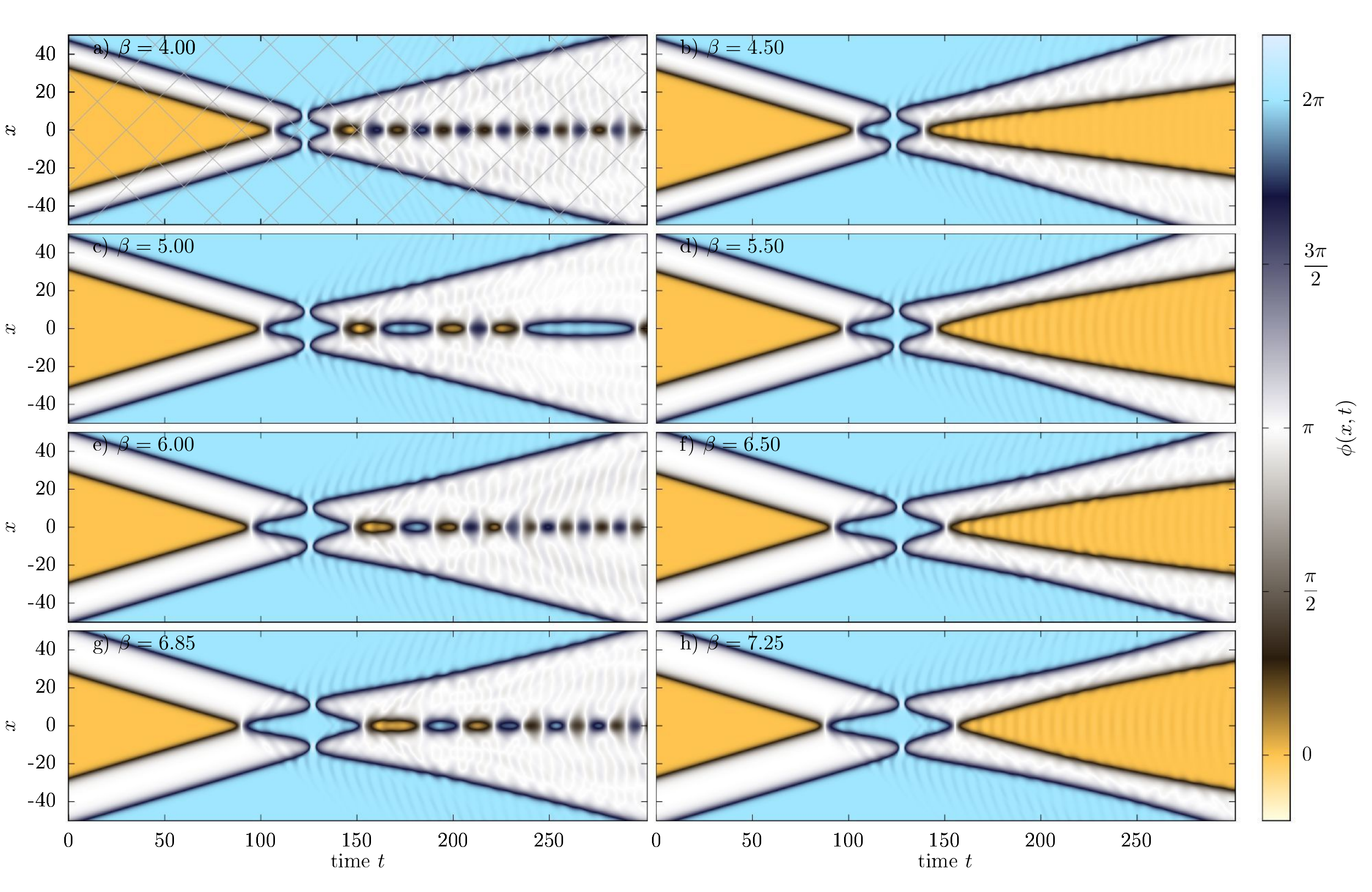}
\end{center}
\vskip -25pt
\caption{Example collisions with initial velocity $v=0.3$ for a variety of values of $\beta$, showing the decomposition of the initial collision into four subkink collisions.}
\label{beta_collection1_2x4}
\end{figure}


\begin{figure}[!h]
\begin{center}
\includegraphics[width=1\textwidth]{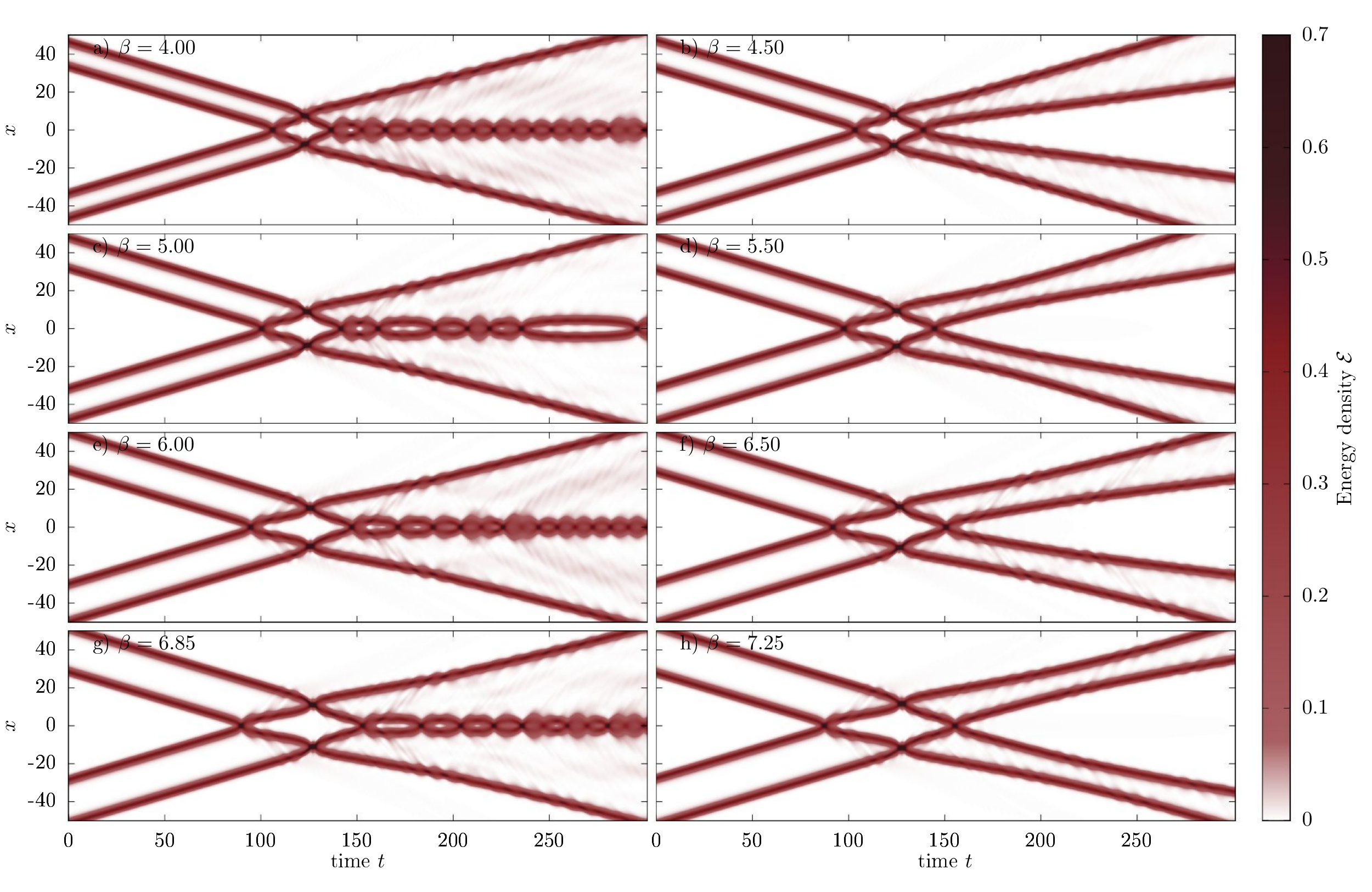}
\end{center}
\vskip -25pt
\caption{Energy density plots for the collisions shown in figure \ref{beta_collection1_2x4}.}
\label{energy_collection1_2x4}
\end{figure}


\begin{figure}[!ht]
\begin{center}
\includegraphics[width=1.0\textwidth]{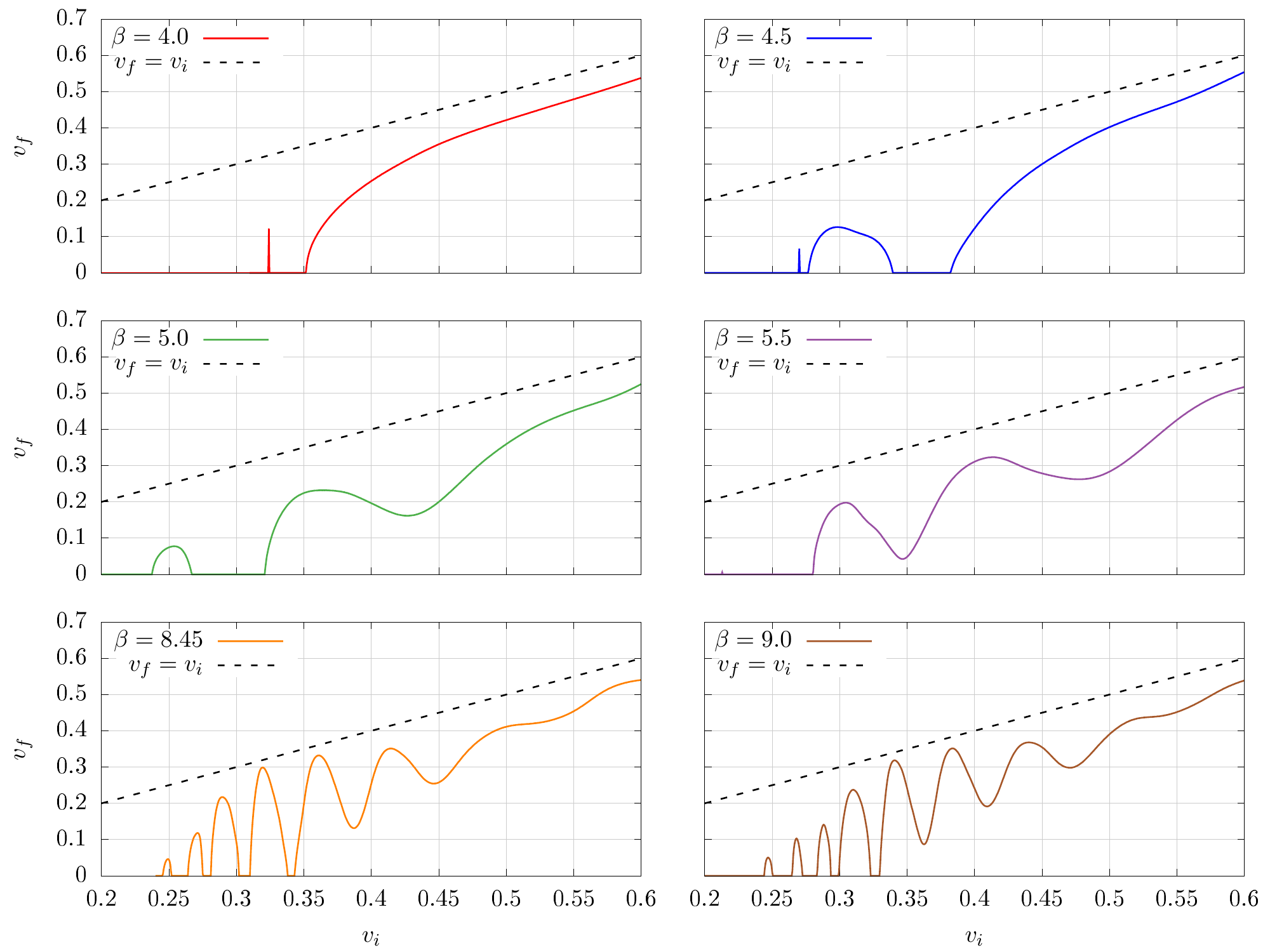}
\end{center}
\vskip -15pt
\caption{Inner subkink velocity shortly after the second subkink -- subkink collision at $x=0$ as a function of initial velocity in the near-critical region, illustrating the mechanism for the formation of the overhanging windows.}
\label{final_velocity}
\end{figure}

\begin{figure}[!ht]
\begin{center}
\includegraphics[width=1\textwidth]{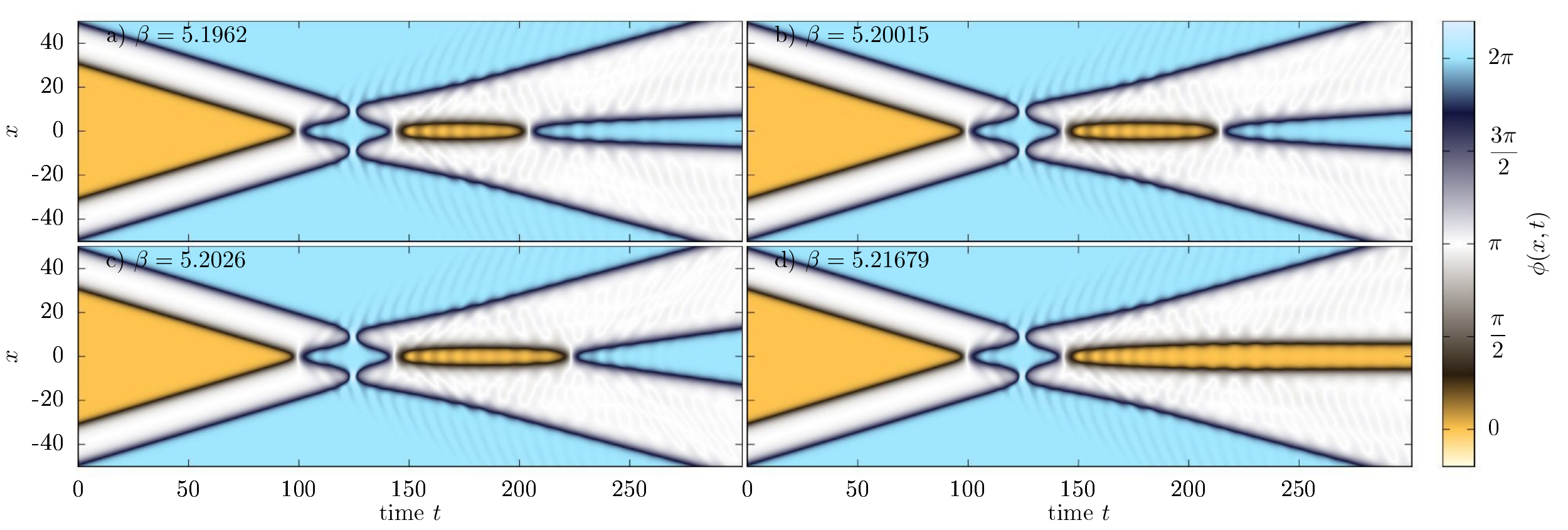}
\end{center}
\vskip -20pt
\caption{Example collisions with initial velocity $v=0.3$ for a variety of values of $\beta$.
In the first three cases
a subsequent resonant collision leads to the formation of a pseudowindow.}
\label{beta_collection1_2x2}
\end{figure}

This surprisingly intricate structure appears to be associated with the dissociation of the kinks into pairs of subkinks in this regime.
Related issues were recently discussed in~\cite{Zhong:2019fub}, but in our model we can tune the distance between the half-kinks by varying the deformation parameter $\epsilon$, and make the double kinks infinitely wide as $\epsilon\to\epsilon_{cr}$.
The initial $\bar KK$ collision no longer happens as a single event, but rather splits into four subcollisions between the weakly bound constituent subkinks, with the fourth collision being key in determining the ultimate fate of the process. 
Some examples are shown in figure \ref{beta_collection1_2x4}, while figure
\ref{energy_collection1_2x4} shows the corresponding energy densities, highlighting the
fragmented nature of the kinks in this regime. 
The first of the four subcollisions results in a bounce of the inner subkinks combined with a vacuum 
flip at the centre. The inner subkinks, moving slightly slower than before, then collide with the 
outer subkinks which arrive slightly later due to the extended structure of the full defect, and bounce 
back again. At least for the processes shown in figure \ref{beta_collection1_2x2}, these 
returning subkinks are moving faster than were the outgoing subkinks that resulted from the first bounce. 
This might be surprising, but note that the outer subkinks arrive at the second 
subcollision travelling at full speed, while the outgoing inner subkinks are moving more slowly, so
the rest frame for the second subcollision is moving towards the centre.   
From this point the scenarios differ depending on the velocity and $\beta$. 
The ingoing subkinks can annihilate at the centre to 
form a bion surrounded by a region of false vacuum, as in the 
processes plotted in the left-hand columns of figures \ref{beta_collection1_2x4} and 
\ref{energy_collection1_2x4}. These processes correspond to points inside
the first four spines shown in figure \ref{nearcritscan_zoom}.
Alternatively,
the inner subkinks can escape from this fourth subcollision
to pair off again with the outer subkinks, as in the processes shown 
in the right-hand columns of figures \ref{beta_collection1_2x4} and 
\ref{energy_collection1_2x4},
with the 
vacuum at the centre flipping back to its original value. These correspond to points 
in the yellow-coloured regions between the spines.

This subkink capture/escape scenario is itself subject to a resonance mechanism, signs of
which can be seen by counting the oscillations in the false vacuum visible between the first and
second subcollisions in the processes shown in the right-hand 
column of figure \ref{beta_collection1_2x4}.
Figure \ref{final_velocity} illustrates aspects of this scenario
in more detail. The final velocity of the inner kink after the second central subcollision reveals some oscillational dependence on the initial velocity. This is very different from the monotonic growth above the upper critical velocity  observed in the case of collisions of unexcited defects. However, exactly this feature was recently observed in \cite{Alonso-Izquierdo:2020ldi} in the case of scattering of wobbling kinks. In that paper it was shown that if the excitation of the kinks is large enough it can lead to the appearance of new windows above the upper critical velocity. This is exactly the same effect that we observe for larger values of $\beta$ with the exception that the excitation is not bound to the subkinks but is a superposition of radiation and inner oscillations of the double kink. Scanning through figure \ref{final_velocity} as $\beta$ varies gives a good intuition of how the spines in figure \ref{nearcritscan_zoom} are formed.

More complicated evolutions after the fourth subcollision are also possible, leading to 
the appearance of windows and pseudowindows within the spines, provided suitable resonance conditions are met. Figure \ref{beta_collection1_2x2} 
shows three examples corresponding to three of the pseudowindows visible in figure
\ref{ScanBeta1D_c}, and one more showing the field evolution just outside the spine shown in that figure.


The presence of the outer subkinks can also disturb the long-term evolution of the system. 
When a central bion is formed, the outer subkinks initially escape from the centre of collision, but because of the presence of the false vacuum they will always return and recollide with the bion at the centre. The time scale for this recollision is large, comparable with the period of the lowest oscillations of the double kink, which is of order $e^\beta$ (see section \ref{sec:frequencies}). The collision between the returning outer subkinks and the bion is a chaotic process depending on many factors including the velocity of the incoming subkinks, the amplitude and the phase of the bion at the moment of the collision, and omnipresent radiation. 

\begin{figure}[!ht]
\begin{center}
\includegraphics[width=1\textwidth]{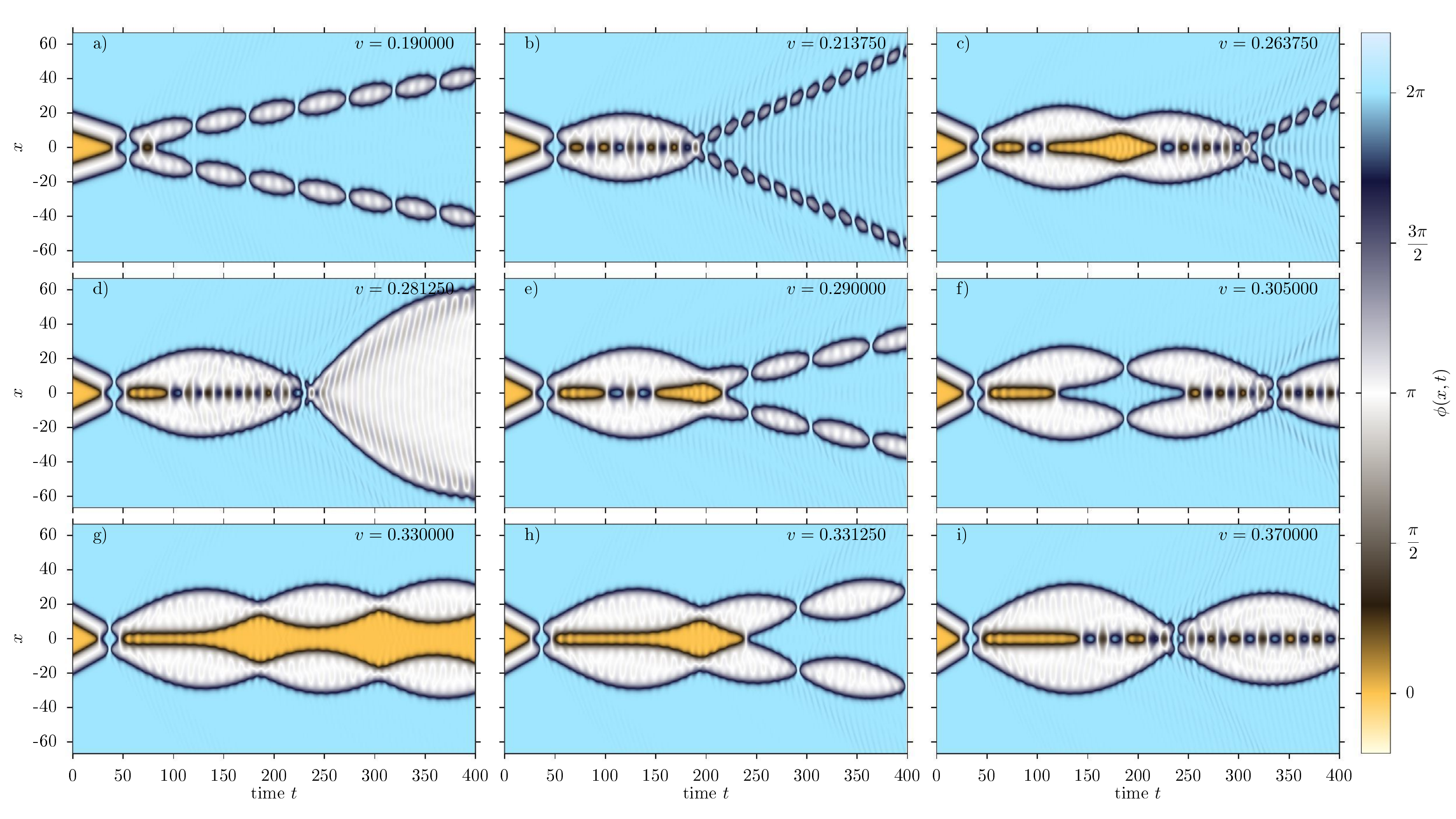}
\end{center}
\vskip -20pt
\caption{Example scattering processes for a variety of initial velocities at $\epsilon=2.59935$ 
($\beta = 3.09887$), illustrating the variety of behaviours of the subkinks after the initial four-way collision.}
\label{nearcriticalmaps}
\end{figure}

Some examples of such collisions are shown in figure \ref{nearcriticalmaps}. One striking feature is that near $\epsilon_{cr}$ large bubbles of the false vacuum can be created (white regions). Such bubbles tend to shrink but the presence of the central bion can impede their vanishing,
as in plots (d), (f) and (i) of figure \ref{nearcriticalmaps}. 
In many cases as a final result two bions are ejected from the collision centre (plots (a), (b), (c) and (h) of figure \ref{nearcriticalmaps}). These events lie in the pseudowindows introduced above, signalled by regions of light blue colour on figures \ref{velocityscanx3} and \ref{nearcritscan}.

\begin{figure}[!ht]
\begin{center}
\includegraphics[width=1\textwidth]{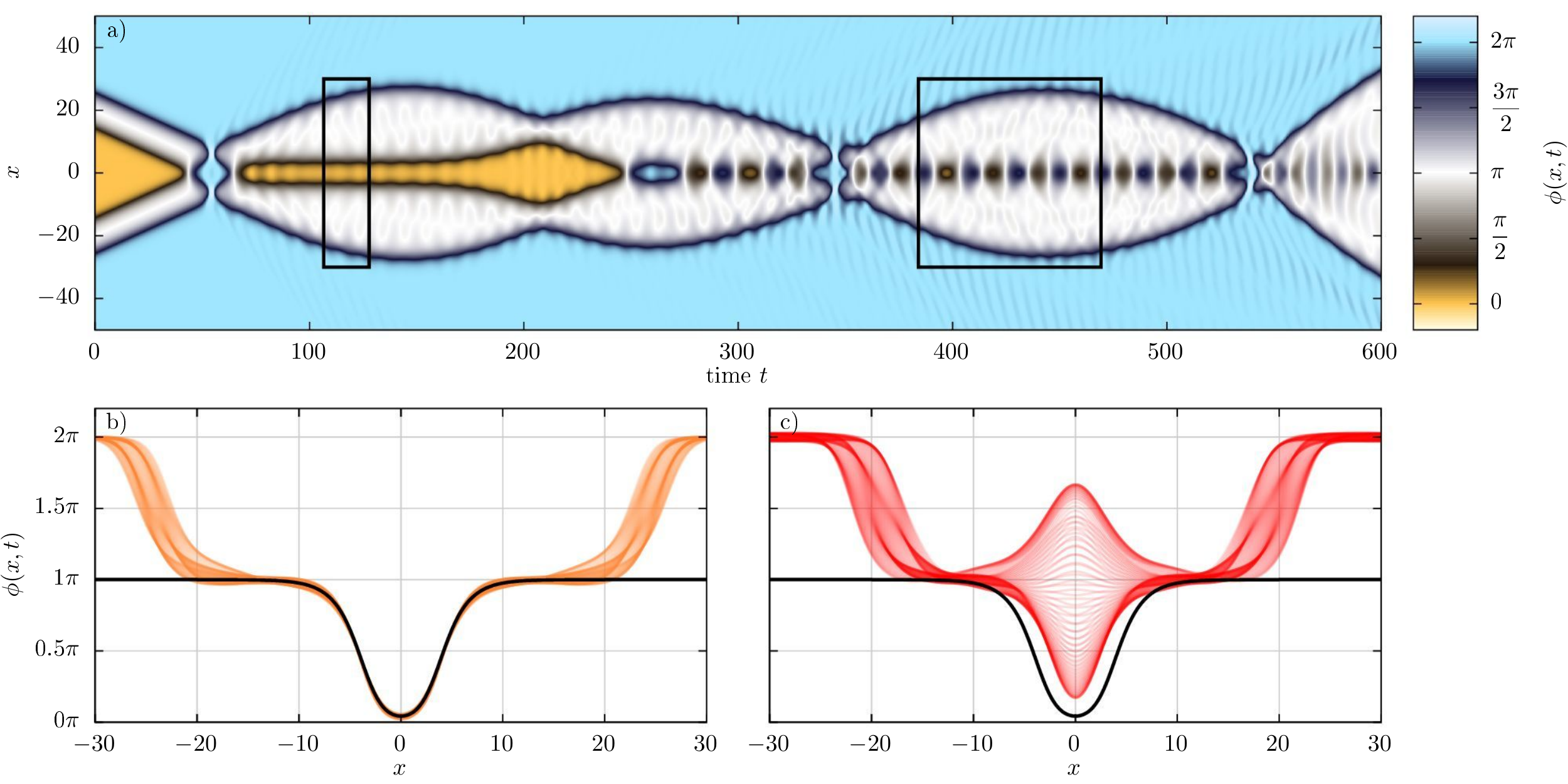}
\end{center}
\vskip -15pt
\caption{A further collision process, comparing the profiles of an intermediate state (b) and a 
bion (c) to the unstable lump solution (black line), for $\epsilon=2.59935$ ($\beta = 3.09887$) and $v=0.31$.}
\label{unstable_dynamics}
\end{figure}

Figure \ref{unstable_dynamics} shows a further example. After the initial collision a false vacuum bubble with an unstable lump (described in section \ref{sec:unstable_lumps}) at centre is created. The
field profile matches the profile of the lump out to $x\approx \pm15$, beyond which
the outer subkinks  connect with the true vacuum, as shown in
figure \ref{unstable_dynamics} (b). At  $t\approx 160$ the unstable lump 
decays, two subkinks
being ejected only to bounce off the outer subkinks and recollide at centre, forming a bion. This bion survives the first collision with the outer subkinks around $t\approx350$. Its profile is encapsulated in the profile of the unstable lump (figure \ref{unstable_dynamics} (c)). 
In a sense the lump can be thought of as a critical profile of a bion with zero frequency. During the second collision at $t\approx640$ the bion is highly perturbed and much of its energy is transferred to the outer subkinks, which clearly move with higher velocity after the bounce.

\begin{figure}[!ht]
\begin{center}
\includegraphics[width=1\textwidth]{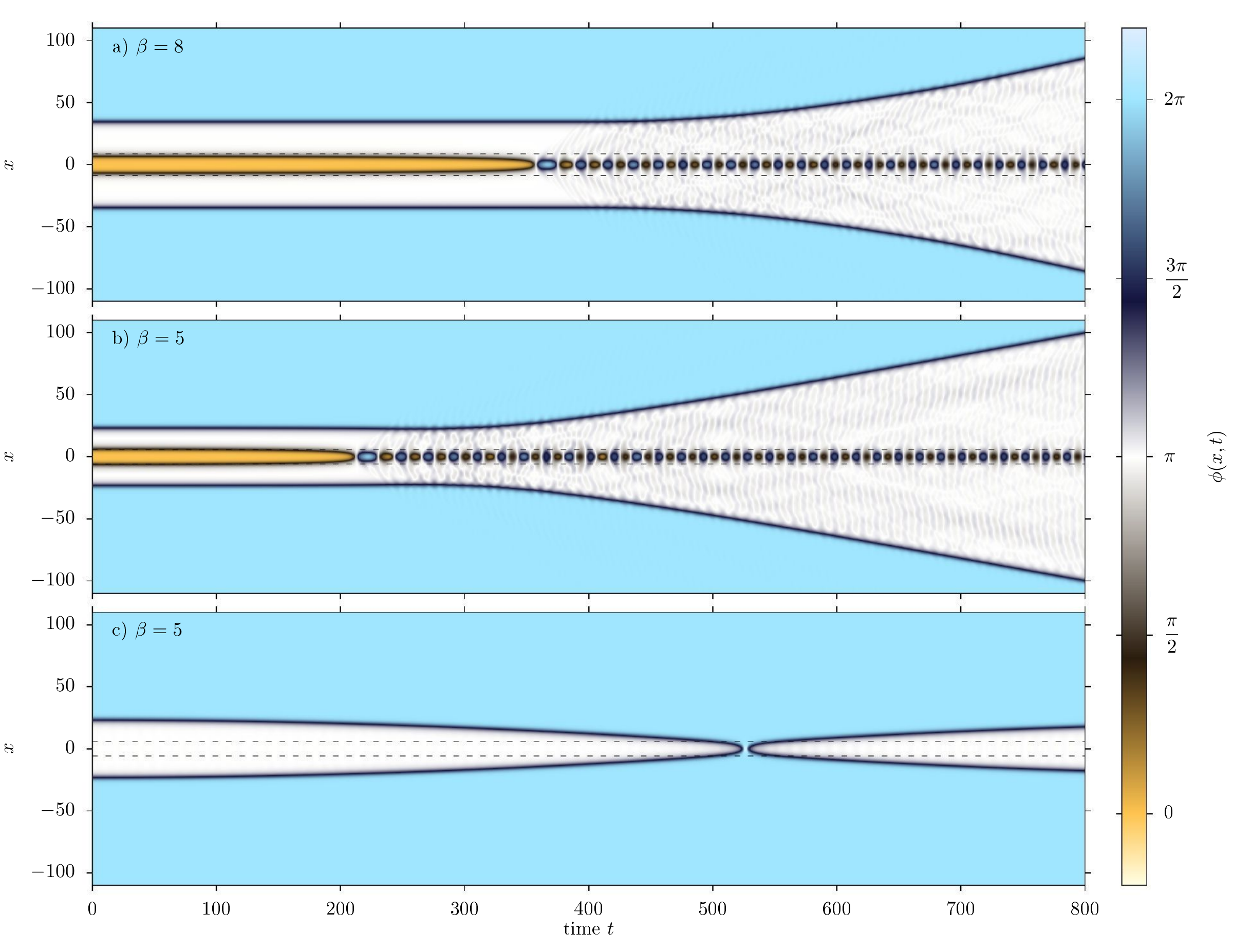}
\end{center}
\vskip -15pt
\caption{``Blowing bubbles": two static kinks placed close to each other, for two values of $\epsilon$ near to $\epsilon_{cr}$. The innermost subkinks attract each other and annihilate to create a bion, the radiation from which pushes the outer subkinks apart, thereby expanding the region of false vacuum. The final plot depicts the field evolution for the second case with the innermost subkinks removed, showing the collapse (and rebound) of the false vacuum bubble in the absence of radiation pressure.}
\label{mapsPressurec}
\end{figure}

For $\epsilon$ very close to $\epsilon_{cr}$, the energy of the false 
vacuum is only very slightly lifted above that of the true
vacuum, and the vacuum pressure (which acts to shrink the region of false vacuum) is small. This means 
that other effects such as radiation pressure can be equally important. In this regime of
our model the vacua on either side of the defect have significantly different
masses associated with their small fluctuations, with the larger mass belonging to the true vacuum. 
As a result, radiation pressure on this defect acts in the opposite direction to vacuum pressure, expanding regions of false 
vacuum.
(Similar radiation pressure effects, albeit in the absence of vacuum pressure, were previously discussed 
in \cite{Romanczukiewicz:2017hdu}.)
Signs of this effect can be seen in the plots in the left-hand column of
figure \ref{beta_collection1_2x4}, and in figure
\ref{beta_collection1_2x2}, in the slight acceleration of the outermost subkinks after the initial four-way collision. To illustrate the phenomenon and its relevance more clearly, the first two plots of
figure \ref{mapsPressurec} show the time-evolution of an initially-static configuration  corresponding to a very closely-placed kink and antikink pair. The innermost subkinks, one from the kink and one from the
antikink, attract each other and annihilate to
create a bion a finite time after the start of the simulation. This bion then acts as a source of radiation embedded in 
a bubble of false vacuum,
causing it to expand. The final plot shows that in the absence of radiation, the bubble instead contracts.

\subsection{Small kinks at critical $\epsilon$}
To finish, we discuss the situation when $\epsilon=\epsilon_{cr}$ 
and our model has three exactly-degenerate vacua $\{0, \pi, 2\pi\}$.
The full kink ceases to exist as a static configuration, and is replaced by two small kinks, $(0,\pi)$ and $(\pi,2\pi)$, which are independent topological solitons, along with the corresponding small antikinks. While
the potential is not quite the same, the scattering of these
small kinks and/or small antikinks is very similar to that of kinks and antikinks in the triply-degenerate $\phi^6$ model.

\begin{figure}[!ht]
\begin{center}
\includegraphics[width=1\textwidth]{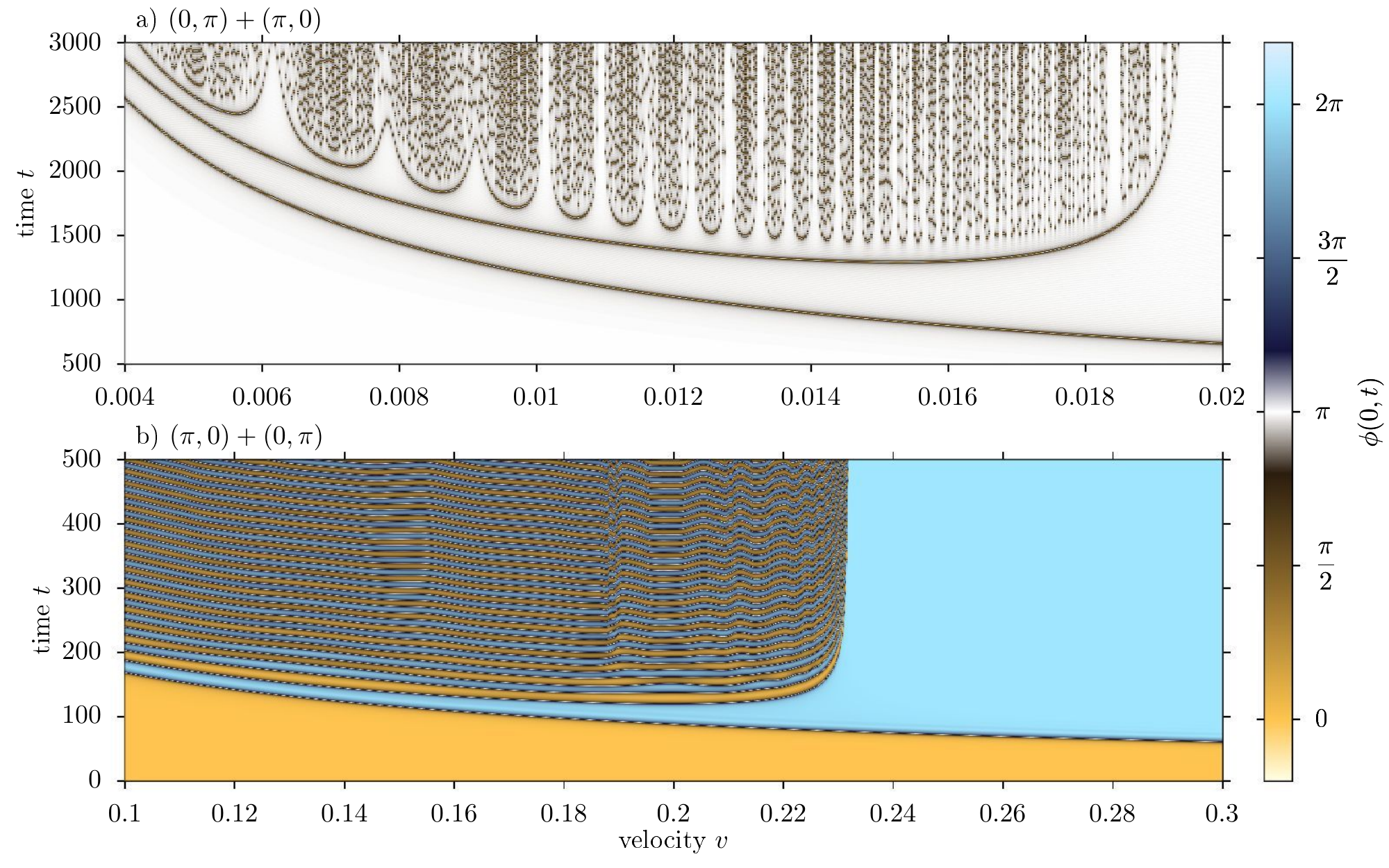}
\end{center}
\vskip -15pt
\caption{Collision of small antikinks and kinks for critical value $\epsilon=\epsilon_{cr}$ for alignments $(0,\pi)+(\pi,0)$ (a) and  $(\pi, 0)+(0, \pi)$ (b).}
\label{halfkinks}
\end{figure}

Two small kinks  repel each other and their scattering shows no resonant phenomena. Two types
of collisions between a small kink and a small antikink are possible, depending on which
vacuum sits between them. The mass of the small perturbations around
the $\phi=\pi$ vacuum is lower than that
 around $\phi=0$ and $\phi=2\pi$. 
 When a $(\pi, 0)$ antikink collides with a $(0, \pi)$ kink, a
potential well for the spectrum of small fluctuations is generated, 
within which many modes can exist. The width of the well depends on the distance between the
defects. In this alignment the collisions do show a resonant structure, of the sort
first discovered for the
$\phi^6$ model in \cite{Dorey:2011yw}. The resonant structure for this case is shown in
figure~\ref{halfkinks} (a).

By contrast, the collision of a $(0,\pi)$ kink to the left and a $(\pi,0)$ antikink to the right
creates a potential bump between the defects where no bound modes can exist, and as a result 
collisions of this type show no resonant structure, as seen in figure~\ref{halfkinks} (b).

The pattern of true and false windows visible in figure~\ref{halfkinks} (a) is similar to that seen previously, for the 
triply-degenerate $\phi^6$ model, in \cite{Dorey:2011yw}. However the colour scheme used
in figure~\ref{halfkinks}  -- chosen for consistency with the one used elsewhere in this paper -- makes the precise structure hard to discern.

\begin{figure}[!ht]
\begin{center}
\includegraphics[width=0.95\textwidth]{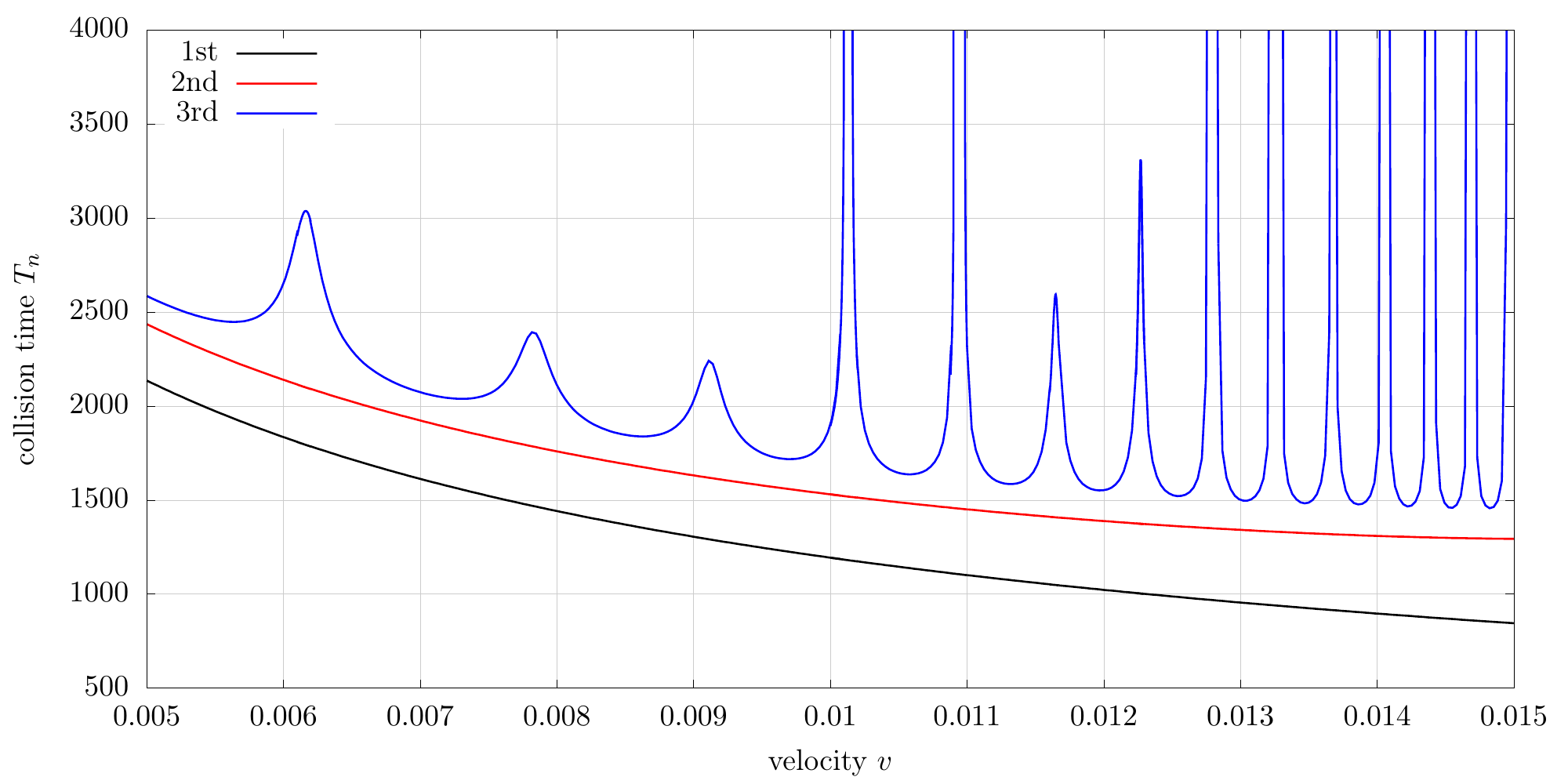}
\end{center}
\vskip -15pt
\caption{A closer look at the processes shown in figure \ref{halfkinks} (a), showing the times to the first, second and third bounces for the scattering of small kinks and antikinks in the $(0,\pi)+(\pi,0)$ configuration
at $\epsilon=\epsilon_{cr}$. }
\label{halfkinktimes}
\end{figure}

Figure \ref{halfkinktimes}, which should be compared with figure 1 (e) of \cite{Dorey:2011yw},  shows a more detailed picture. Similar to the situation for the $\phi^6$ model, but different from the sequences of closing windows seen earlier in this paper, we see that true and false windows are interleaved, something which might be associated with the fact that multiple resonant frequencies play a role in 
these cases. However the patterns are slightly different -- for the $\phi^6$ model there is a single isolated false window directly following the first true window, while here two true windows are followed by two false ones. An interesting challenge to any effective model of kink -- antikink collisions in models of this type, where the resonant modes live in the gap between the colliding solitons, would be to reproduce this distinction.

\section{Conclusions}

In this paper we have studied the dynamical properties
of a scalar field theory interpolating between the 
completely integrable sine-Gordon model, the usual $\phi^4$ theory, and a $\phi^6$-like theory with three 
degenerate vacua.
Our numerical simulations of the collisions between kinks and antikinks in the model revealed a rich
diversity of behaviours and suggest a number of avenues for further investigation.

For $\epsilon\ll 1$ the key feature of the model is the breaking of integrability. We identified the role of the false 
vacuum in the smooth transition from the kink - antikink reflection that occurs for $\epsilon>0$ to the transmission
found
at $\epsilon=0$. Associated with this is a sharp decrease in the critical velocity $v_{cr}(\epsilon)$, down to zero at $\epsilon=0$ where integrability is restored. This is clear from our numerical results, but an analytic
understanding of the behaviour of $v_{c}(\epsilon)$ for $\epsilon\ll 1$ would be very valuable. This regime would also be a good starting-point for an investigation of the quantum theory of this model.

In the
neighbourhood of $\epsilon=1$, we saw a smooth deformation of the $\phi^4$ picture; again we
would like to understand the dependence of $v_{c}(\epsilon)$ (and indeed the window locations) on $\epsilon$ from
an analytic point of view. At both ends of this regime, resonance windows transition from true to false, starting 
with those at the lowest velocities. As things stand this is a numerical observation.

Beyond $\epsilon\approx 1.5$ the structure of the resonance windows becomes richer and more chaotic, with the emergence of
novel structures that we called pseudowindows. In spite of the increased complexity of the behaviour of the model in this regime, we did see some signs of regularity, as reported in table \ref{extrema}.

The region $\epsilon\to\epsilon_{cr}$ is particularly intriguing. Our zoomed-in scans, figures 
\ref{nearcritscan} and \ref{nearcritscan_zoom}, revealed a 
novel and unexpected structure of `spines', which we associated with the emergence of an internal 
structure to our kinks in this limit. 
We identified some of the relevant mechanisms, in particular in the decomposition of individual kink - antikink collisions into
four subcollisions, but more work is needed to pin down the details. Preliminary 
studies show that similar structures arise in the somewhat-simpler model studied by Christ and Lee in
\cite{Christ-Lee}. Numerical work in this regime is particularly delicate as the presence of the false vacuum 
means that some relevant processes take place over extremely long timescales, and whether greater regularity will emerge 
in the extreme limit $\epsilon\to\epsilon_{cr}$, $\beta\to\infty$ remains an open question; but these same effects make the model in this regime an excellent arena for the study of phenomena such as radiation pressure.
In this region we also observed the formation of bions/oscillons submerged in a bubble of false vacuum. Their shapes were bounded by the profile of an unstable lump, 
a static solution consisting of two subkinks placed in unstable equilibrium, balancing false vacuum pressure against  the mutual attraction between subkinks.

Finally, exactly at the limiting value $\epsilon=\epsilon_{cr}$ a triply-degenerate vacuum structure similar to
the $\phi^6$ model was restored. The subkinks become fully independent defects and their collisions, characteristic of non-symmetric kinks, happen with very similar mechanisms to those found in the $\phi^6$ model. 
In spite of these similarities, the structure of false and true windows was different. 

To conclude, kink scattering in the deformed sine-Gordon model exhibits a remarkable variety of phenomena. Some of these we have analysed in detail, while others we have merely pointed out, and merit further study in their own right.
We also expect that many of these newly found effects will be observed in other models, but we will leave the  investigation of this question for future work.

 \acknowledgments
YS gratefully acknowledges partial support of the Ministry of Education  of Russian Federation, 
project  FEWF-2020-0003. 
TR wishes to thank National Science Centre, grant number 2019/35/B/ST2/00059.
PED is grateful for support from the European Union's Horizon 2020 research and innovation programme under the Marie Skodowska-Curie grant agreement No.\ 764850, SAGEX, and from the STFC under consolidated grant ST/T000708/1 “Particles, Fields and Spacetime”.

\bibliographystyle{JHEP}
\bibliography{ref}
\end{document}